\documentclass{article}
\usepackage[T1]{fontenc}
\usepackage{euler}
\usepackage[mathscr]{euscript}
\usepackage{amsmath}
\usepackage{comment}
\usepackage{amssymb}
\usepackage{physics}
\usepackage{tensor}
\usepackage{mathtools}
\usepackage{slashed}
\usepackage{hyperref}
\hypersetup{ 
	colorlinks=true,
	linkcolor=blue,
	urlcolor=blue,
	citecolor=blue
}
\usepackage{geometry}
\usepackage{xcolor}
\usepackage{multirow}
\usepackage{biblatex} 
\usepackage{authblk}
\title{Stable Envelopes, Vortex Moduli Spaces and Verma Modules}
\author{Spencer Tamagni}
\affil{\textit{Center for Theoretical Physics, University of California, Berkeley}}
\date{}

\begin{document}
\setcounter{tocdepth}{2}
\maketitle
\tableofcontents
\section{Introduction}
\subsection{Overview}
The Knizhnik-Zamolodchikov (KZ) equation and its quantum Knizhnik-Zamolodchikov (qKZ) counterpart are equations central to modern mathematical physics. The equations originated from the study of two-dimensional rational conformal field theory (CFT) and integrable lattice models, respectively. They have a multitude of applications, ranging from knot theory \cite{w89} to geometric Langlands duality \cite{frenkel05}.

Recently it was understood that the equations and their solutions, together with the many associated structures originate in superstring theory. This provides a conceptual home for the many recently discovered connections between the equations and geometry and physics, as well as means of discovering new ones. In particular, it provides a framework \cite{a20, a21} for categorification of quantum link and three manifold invariants \cite{gm19, park20} which arise as their monodromy matrices. 

In this paper, guided by string theory, we extend this to a larger class of representations of the underlying (quantum) affine algebras: from those labeled by finite dimensional representations of the corresponding Lie algebra studied in \cite{afo17}, to Verma module representations. We will also describe the most straightforward application of our results, to (a deformed version of) geometric Langlands correspondence with ramifications. While the general picture is motivated by string theory, in the main body of the paper we distill this into a precise mathematical picture and demonstrate all the main claims explicitly in the $\mathfrak{g} = \mathfrak{sl}_2$ case. 
\subsubsection{}
Recall the (trigonometric) KZ equation solved by conformal blocks of $\widehat{\mathfrak{g}}$ on ${\mathbb C}^{\times}$:
\begin{equation}
\kappa a_i\pdv{}{a_i} \psi = \sum_{j (\neq i)} r_{ij}(a_i/a_j)\psi. 
\end{equation}
Its solution $\psi$ takes values in a subspace of a tensor product of evaluation $\widehat{\mathfrak{g}}$-modules associated to the punctures on $\mathbb{C}^\times$ (this includes $0$ and $\infty$):
\begin{equation}
\Big( \bigotimes_i V_i(a_i) \Big)^{\mathfrak{g}}.
\end{equation}
The parameter $\kappa \in \mathbb{C}$ is the level; the $r_{ij}(a_i/a_j)$ matrices act in the $i$-th and $j$-th tensor factors and solve the classical Yang-Baxter equation. The KZ equation has a natural $q$ deformation (the notation will be fully explained in the main body):
\begin{equation}
\begin{split}
\psi(a_1, \dots, qa_i, \dots, a_n) = &  R_{ii-1}(qa_i/a_{i-1}) \dots R_{i1}(qa_i/a_1) (z^h)_i \\ & \times R_{in}(a_i/a_n) \dots R_{ii+1}(a_i/a_{i + 1})\psi(a_1, \dots, a_n). 
\end{split}
\end{equation}
discovered by I. Frenkel and N. Reshetikhin in \cite{frenkelreshetikhin}, where now $R$ is the $R$-matrix of quantum affine algebra $\mathscr{U}_\hbar(\widehat{\mathfrak{g}})$, where $\hbar = q^{-1/\kappa}$, and $(z^h)_i$ is a certain diagonal matrix acting in the $i$-th tensor factor. The $R_{ij}(a_i/a_j)$ matrices solve the quantum Yang-Baxter equation, and reduce to the classical $R$-matrices in the $\hbar\rightarrow 1$ limit. Solutions to the qKZ equation can be viewed as correlation functions of vertex operators of $\mathscr{U}_\hbar(\widehat{\mathfrak{g}})$, viewed as a $q$-deformation of $\widehat{\mathfrak{g}}$. The $q$-deformation breaks conformal invariance. In particular, had we started with the rational form of the KZ equation, associated with conformal blocks on $\mathbb{C}$, we would have discovered a different form of qKZ associated with the Yangian $Y_\hbar(\mathfrak{g})$. 
\subsubsection{}
The fact that it is possible to break conformal invariance while preserving associated structures is remarkable. A physical explanation for this fact may be found in \cite{afo17}. If ${\mathfrak g}$ is a simply laced Lie algebra, one can interpret the deformed conformal blocks of $\mathscr{U}_\hbar(\widehat{\mathfrak{g}})$ as partition functions of a certain six dimensional string theory with $(2, 0)$ supersymmetry and labeled by ${\mathfrak g}$, called little string theory, in the presence of defects. 

The little string theory has a conformal limit in which it becomes the six dimensional $(2, 0)$ superconformal field theory of type ${\mathfrak{g}}$. The conformal limit of the theory turns out to be the limit in which quantum affine algebra $\mathscr{U}_\hbar(\widehat{\mathfrak{g}})$ becomes the  affine Lie algebra $\widehat{\mathfrak{g}}$.

\subsubsection{}
The partition function of little string theory in such a situation can be determined explicitly, thanks to the fact \cite{ah15} that the six dimensional theory localizes to a quantum field theory supported on its defects. Vertex operators labeled by finite dimensional representations of ${\mathfrak{g}}$, studied in \cite{afo17, a20, a21}, lead to defect theories which are three-dimensional quiver gauge theories with $\mathscr{N} = 4$ supersymmetry compactified on a circle. The quivers are based on the Dynkin diagram  of ${\mathfrak{g}}$. The relevant supersymmetric partition functions are controlled by $K$-theoretic enumerative geometry of Nakajima quiver varieties (the Nakajima variety in question is the Higgs branch of vacua of the gauge theory). The identification of the $q$-difference equations controlling the $K$-theoretic curve counts with qKZ is a mathematical result due to \cite{okounkov15, os16, ao17}. In the conformal limit, the theories on the defects of the $(2,0)$ SCFT become strongly coupled, and generically are no longer gauge theories. Their partition functions are computed via a careful $\hbar \to 1$ limit of the relevant gauge theory calculation \cite{afo17}. 

Our main result is an extension of this to conformal blocks of $\mathscr{U}_\hbar(\widehat{\mathfrak{g}})$ with insertions of Verma module vertex operators at \textit{generic complex highest weight}. From the little string perspective, this corresponds to including codimension two defects originally studied in \cite{ah15}, as opposed to the codimension four defects of \cite{afo17}. The theories on the defects are three-dimensional gauge theories with only $\mathscr{N} = 2$ supersymmetry. 
Despite the reduced supersymmetry, the resulting theories belong to a distinguished class which share many of the properties of their $\mathscr{N} = 4$ analogs -- in particular, the connection to (geometric) representation theory.
 \subsubsection{}
From the mathematical perspective, the reduced supersymmetry requires us to extend certain results in enumerative geometry of Nakajima varieties \cite{okounkov15} to a more general class of varieties which are not generically holomorphic symplectic. The varieties under study here can be viewed as certain moduli spaces of quasimaps, known to physicists as vortices.  We will explicitly prove all of our claims in the simplest $\mathfrak{g} = \mathfrak{sl}_2$ case, and make precise conjectures for a general simply laced $\mathfrak{g}$.
Already in the $\mathfrak{sl}_2$ case there are quite a few technical details on the geometric side that must be addressed. Stable envelopes in K-theory and elliptic cohomology are the key tool in our construction; this is to our knowledge one of the first concrete applications of the recent wider class of existence results \cite{okounkov21, okounkov20} (see \cite{zhouetal2024} for another application in the context of superspin chains). 
\subsubsection{}
We also provide an application of our results to a deformed version of quantum geometric Langlands correspondence. The correspondence, proposed in \cite{afo17}, relates $q$-deformed conformal blocks of $\mathscr{U}_\hbar(\widehat{^L\mathfrak{g}})$ and $\mathscr{W}_{q, t}(\mathfrak{g})$ where $^L\mathfrak{g}$ is the Lie algebra Langlands dual to $\mathfrak{g}$. We introduce the ramified version of this correspondence, explain its string theory origin, and verify it in the simplest case. For a much more extensive general discussion of ramified quantum $q$-Langlands correspondence see \cite{haouzi23}.

Our results may also be applied to the knot categorification program as in \cite{a20, a21}; in this case, the invariant to categorify is the conjectural $\widehat{Z}$ invariant of three-manifolds, specialized to three-manifolds which are knot complements. A mathematical definition was proposed, at least for a large class of knots, in \cite{park20} using the braiding matrices associated to Verma module vertex operators of $\widehat{\mathfrak{g}}$, which provides the basic link to the conformal $\hbar \to 1$ limit of the constructions in this paper. The categorification of this construction along the lines of \cite{a21} will be discussed in a separate publication \cite{at23}. 
\subsubsection{}
Because the $\mathfrak{g} = \mathfrak{sl}_2$ case is already interesting for applications, we do not go beyond this case in this paper as far as mathematical proofs are concerned. The string theory origin of our results predicts the generalization to arbitrary ADE Lie algebras $\mathfrak{g}$ which we spell out as a conjecture. The general picture and string theory background will be sketched in section \ref{generalg}. Certainly it would be of interest to put the broader picture on similarly firm mathematical ground, but we leave this work to others.
\subsubsection{}
In what follows, we outline our results in greater detail in the $\mathfrak{g} = \mathfrak{sl}_2$  case, provide some elaboration on the mathematical point of view, and point out connections to other work that we are aware of.  

\subsection{Geometry and representation theory of $\mathscr{U}_\hbar(\widehat{\mathfrak{sl}_2})$ }

There is a well known relation of
\begin{equation}
Y(k,n)= T^*\text{Gr}(k, n),
\end{equation}
the cotangent bundle to the Grassmannian of $k$-planes in $\mathbb{C}^n$, to representation theory of $\mathfrak{sl}_2$ and its quantum group variants. As we will review below, the choice of $k$ and $n$ is related to picking a subspace of fixed weight, $k$ levels below the highest, inside the the tensor product of $n$ copies $V_{1/2}^{\otimes n}$ of the fundamental (spin-$1/2$) representation of $\mathfrak{sl}_2$.

The main actor in this paper is a space 
\begin{equation}\label{Xkn}
 X(k, n)
\end{equation}
which corresponds to 
replacing the $n$ copies of the two-dimensional  $V_{1/2}$ representation by $n$ copies of Verma module representations with arbitrary complex highest weights $\mu_1,\ldots, \mu_n$, and again fixing a weight subspace $k$ levels below the maximal.
\subsubsection{} 
The space $Y=Y(k,n)$ is a Nakajima quiver variety, the Higgs branch of an $A_1$ quiver gauge theory with $\mathscr{N}=4$ supersymmetry in three dimensions. The torus $T = A \times \mathbb{C}^\times_\hbar$ acts on $T^*\text{Gr}(k, n)$, where $A \subset GL_n$ is a maximal torus and $\mathbb{C}^\times_\hbar$ acts by scaling the cotangent fibers. The equivariant parameters associated to $A$ are denoted $(a_1, \dots, a_n)$ and the equivariant parameter associated to $\mathbb{C}^\times_\hbar$ is denoted $\hbar$. There is an action of $\mathscr{U}_\hbar(\widehat{\mathfrak{sl}_2})$ on the $T$-equivariant K-theory of $\bigsqcup_{k = 0}^n T^*\text{Gr}(k, n)$ which produces an identification
\begin{equation}
K_T \Big(T^*\text{Gr}(k, n) \Big)\otimes_{K_T(\text{pt})} (\text{field}) = \Bigg( \mathbb{C}^2(a_1) \otimes \dots \otimes \mathbb{C}^2(a_n) \Bigg)_{\text{weight} = \frac{n}{2} - k}.
\end{equation}
The right hand side denotes the subspace of weight $\frac{n}{2} - k$ in a tensor product of $n$ fundamental evaluation modules of $\mathscr{U}_\hbar(\widehat{\mathfrak{sl}_2})$ with evaluation parameters $a_1, \dots, a_n$. This identification can be constructed as the simplest application of the apparatus of K-theoretic stable envelopes for Nakajima quiver varieties \cite{okounkov15}, \cite{os16}. The product on the left hand side just means that we work over the fraction field of $K_T(\text{pt})$, so equivariant parameters for $T$ are considered fixed at some generic background values. 

The deformed conformal blocks of $\mathscr{U}_\hbar(\widehat{\mathfrak{sl}_2})$ arise in the geometric context as K-theoretic vertex functions of $Y$. The vertex function is a certain generating function of $K$-theoretic counts of curves $\mathbb{P}^1 \to T^*\text{Gr}(k, n)$, which arises in physics by computing the path integral of the underlying 3d $\mathscr{N} = 4$ gauge theory on $\mathbb{C} \times S^1$ by supersymmetric localization. The path integral is computed in the three-dimensional $\Omega$-background, so there is a certain twist by rotation, flavor, and $R$-symmetries when going around the $S^1$, which we will review below. These vertex functions satisfy $q$-difference equations which can be rigorously identified with the qKZ equations acting on a fixed weight subspace of $\mathbb{C}^2(a_1) \otimes \dots \otimes \mathbb{C}^2(a_n)$. $K$-theoretic stable envelopes provide a particular basis of insertions at $0 \in \mathbb{C}$ which conjugate these $q$-difference equations to the standard form, see \cite{ao17}. The parameter $q$ of the difference equations is identified with the automorphisms of the base curve which is mapped to $T^*\text{Gr}(k, n)$. 

Since the naive space of maps $\mathbb{P}^1 \to T^*\text{Gr}(k, n)$ is not compact, to compute the partition function, one must choose a specific compactification. In this case, a natural compactification is provided by the gauge theory itself, and is known as quasimaps (see \cite{okounkov15} for an introduction; we also recall basics on quasimaps in the main body). The compactification comes from gauge theory because singularities of degenerate maps $\mathbb{P}^1 \to T^* \text{Gr}$ are simply the appearance at low energies of underlying smooth vortex solutions in the gauge theory defined in the ultraviolet \cite{lns2000}. 
\subsubsection{}
The space $X(k,n)$ in \eqref{Xkn} is itself a space of quasimaps:
\begin{equation}
X(k, n) = \text{Moduli space of degree $k$ quasimaps $\mathbb{P}^1 \to T^*\text{Gr}(n, 2n)$, based at $\infty$}. 
\end{equation}
It is a smooth quasiprojective variety of dimension $2kn$. A torus $T$ of rank $2n + 1$ acts on $X(k, n)$.  The equivariant parameter $\hbar$ in this case is associated to automorphisms of the domain curve.  Unlike $Y(k,n)$, it is not a Nakajima quiver variety; rather $X(k,n)$ is a Higgs branch of an $A_1$ quiver gauge theory with only $\mathscr{N}=2$ supersymmetry in three dimensions. It admits a description as a so-called handsaw quiver variety \cite{nak11}. This description in particular implies the existence of stable envelopes for $X(k, n)$ from Theorem 1 of \cite{okounkov20}. We explicitly write down the stable envelopes in the main body. 

A byproduct of the existence of stable envelopes (and explicit formulas for them) is an action of $\mathscr{U}_\hbar(\widehat{\mathfrak{sl}_2})$ on equivariant $K$-theory, and an identification 
\begin{equation} \label{rep}
K_T\Big( X(k, n) \Big) \otimes_{K_T(\text{pt})} (\text{field}) = \Bigg( \mathscr{V}_{\mu_1}(a_1) \otimes \dots \otimes \mathscr{V}_{\mu_n}(a_n) \Bigg)_{\text{weight} = \frac{1}{2} \sum_\alpha \mu_\alpha - k}.
\end{equation}
$\mathscr{V}_{\mu_\alpha}(a_\alpha)$ denotes an evaluation Verma module of $\mathscr{U}_\hbar(\widehat{\mathfrak{sl}_2})$, with evaluation parameters $a_\alpha$ and highest weight $\mu_\alpha$. The $2n$ parameters $(a_\alpha, \mu_\alpha)$ comprise the rest of the equivariant variables associated to $T$, and are associated to the maximal torus of $GL_n \times GL_n$ acting on $X(k, n)$ induced by the action on the target $T^*\text{Gr}(n, 2n)$ preserving the basepoint at $\infty \in \mathbb{P}^1$.

We prove that the vertex function of $X(k, n)$ (which we define in a mathematically precise way in the main body) satisfies a $q$-difference equation which is identified with qKZ for $\mathscr{U}_\hbar(\widehat{\mathfrak{sl}_2})$ acting in the right hand side of \eqref{rep}, that is, for $q$-conformal blocks on $\mathbb{C}^\times$ with Verma module vertex operators inserted at the points $a_\alpha \in \mathbb{C}^\times$. The vertex function can be given by a contour integral of Mellin-Barnes type, so we rediscover certain integral solutions to qKZ (see \cite{ao17} and references therein for more on the literature surrounding these integral solutions). The K-theoretic stable envelopes play the same role here as they did in \cite{ao17}, in particular they are closely related to off-shell Bethe eigenfunctions for a dual spin chain. In addition, we construct the elliptic stable envelopes for $X(k, n)$, show that they control the monodromy of $q$-difference equations, and obtain in this way a geometric action of elliptic quantum group in tensor products of Verma modules. This gives a geometric origin for the Verma module conformal blocks of $\mathscr{U}_\hbar(\widehat{\mathfrak{sl}_2})$. 

\subsubsection{}
The stationary $q \to 1$ limit of the qKZ equation recovers algebraic Bethe ansatz for an $n$-site XXZ spin chain with each site supporting a complex spin $\mu_\alpha$. The correspondence of the latter with gauge theory has been known for a long time in the physics literature \cite{ns09}, \cite{Chen_2011}. Here we have put the non-stationary result on solid ground in enumerative geometry. More recent work in the physics literature includes the realization of stable envelopes as interfaces in gauge theories \cite{nekrasovdedushenko1, Bullimore_2022}. We will not use that perspective here, though our constructions certainly fit into this general circle of ideas.

In the $n = 1$ case, the identification \eqref{rep} should be closely related with the specialization of the older results of \cite{tsymb09}, \cite{feigin08} to $\mathfrak{sl}_2$. See also the recent work \cite{rouquier}, which uses a similar geometric setup to construct 2-representations of $\mathfrak{sl}_2$.  However, the philosophy here is rather different, using stable envelopes and focusing on $R$-matrices in the style of \cite{mo12}. 

\subsection{Plan of the paper}
The rest of this work is organized as follows. In section \ref{quivergauge} we consider in detail the three-dimensional $\mathscr{N} = 2$ theory supporting the quasimap space $X(k, n)$ as its Higgs branch. We recall some basics on the vacua and twisted chiral ring of the model, and explain in detail the localization calculation of the partition function. Here we provide precise mathematical definitions of what is computed to make contact with enumerative geometry. We prove a certain integral formula for the vertex function of $X(k, n)$ there, which is well-known in the physics literature in the guise of a 3d $\mathscr{N} = 2$ index. In section \ref{stabs} we discuss the K-theoretic and elliptic stable envelopes, and use them to produce solutions to qKZ from geometry. In section \ref{generalg}, we use string theory to provide a conjectural generalization of our results to arbitrary ${\mathfrak{g}}.$ In section \ref{qlanglands}, we discuss applications to the ramified quantum $q$-Langlands correspondence for $\mathfrak{g} = \mathfrak{sl}_2$. 

The main application of the geometric technology developed in this paper is to the ramified quantum $q$-Langlands correspondence, which is discussed in much greater detail in the work \cite{haouzi23}. We encourage readers to look at that paper for a complementary physical perspective on some of what is discussed here, and a much more detailed overview of the string theory construction giving rise to these mathematical assertions for general $\mathfrak{g}$. 

Finally, we close the introduction with the comment that this is a physics paper and we have accordingly presented the material in a way we believe should be accessible to most physicists. However, our results can be formulated precisely in enumerative geometry and at the same time have a very explicit realization via certain integral formulas. The explicit computations performed in the main body amount to essentially rigorous mathematical proofs of the assertions made above for $\mathfrak{g} = \mathfrak{sl}_2$, though we present them in an informal style more common in the physics literature. 

\subsection{Acknowledgements}
I would like to thank my advisor Mina Aganagic for many useful discussions, support, encouragement, comments on the draft, as well as collaboration on the related work to appear \cite{at23}. I am grateful to Andrei Okounkov for many explanations on enumerative geometry and stable envelopes, and useful feedback in the early days of this project. I also benefited from discussions with Sam DeHority and Che Shen on related topics. I would also like to thank Nathan Haouzi for kindly agreeing to coordinate the release of \cite{haouzi23} with this paper. 

\section{A 3d $\mathscr{N} = 2$ Quiver Gauge Theory} \label{quivergauge}
\begin{figure}
    \centering
    \includegraphics[scale=0.1]{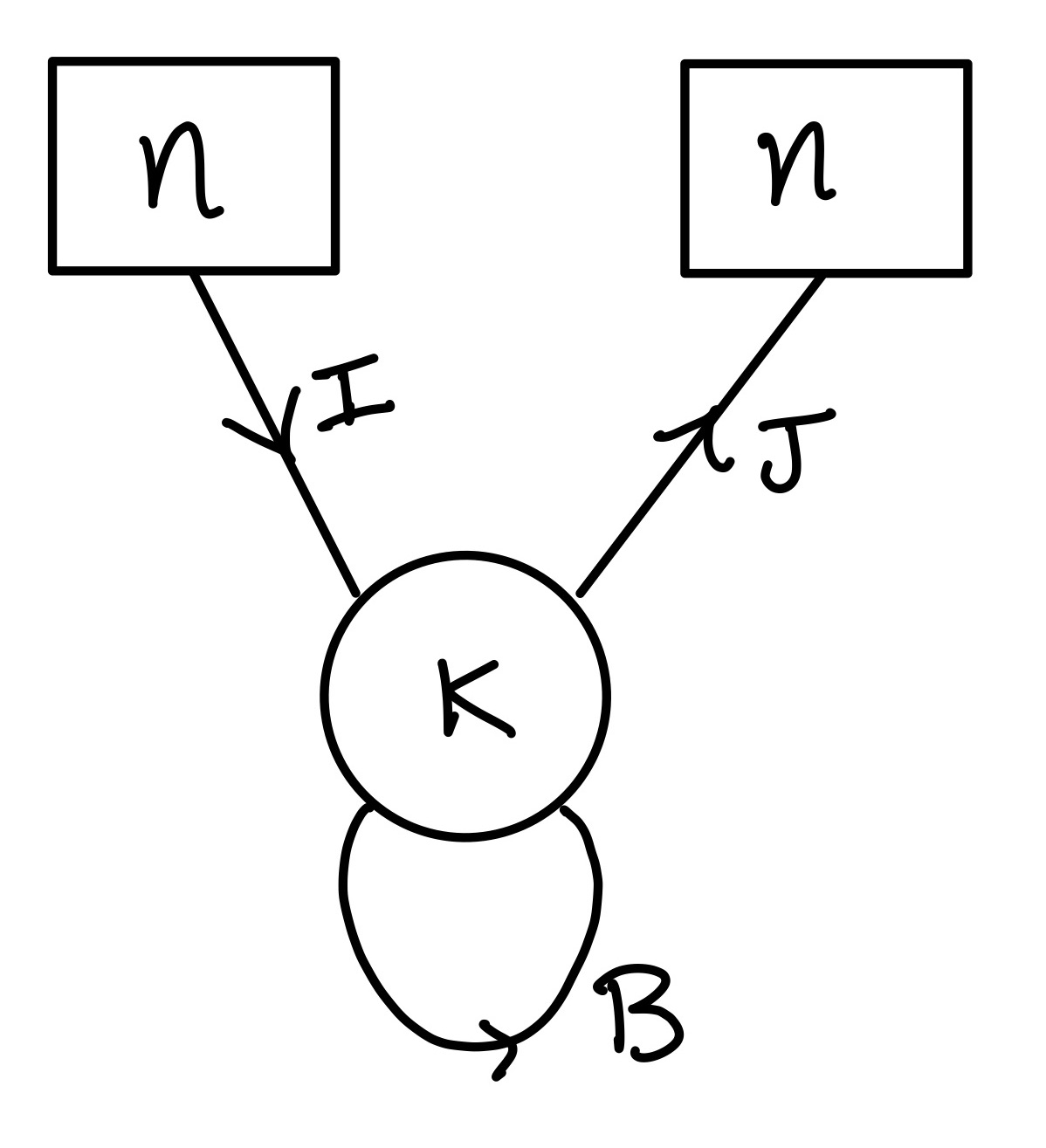}
    \caption{Quiver describing the $\mathscr{N} = 2$ gauge theory.}
    \label{fig:quiver}
\end{figure}

The space $X(k, n)$ in \eqref{Xkn} is the Higgs branch of a handsaw quiver gauge theory with $\mathscr{N} = 2$ supersymmetry (see figure \ref{fig:quiver}). 

The single node of the $A_1$ quiver corresponds to the $U(k)$ gauge group. The loop attached to the gauge node labels a chiral multiplet in the adjoint representation of $U(k)$. The other arrows label $n$ chiral multiplets in the fundamental representation of $U(k)$, and $n$ chiral multiplets in the anti-fundamental representation. Associated to the framing spaces of the handsaw is the group of flavor symmetries 
\begin{equation}
G_F = U(n) \times U(n). 
\end{equation}
The distinguished $U(1)_\hbar$ flavor symmetry acts by scaling the adjoint chiral, and does not act on multiplets containing $I$ and $J$. It should be thought of as inherited from a symmetry of an $\mathscr{N} =4$ theory, obtained by forgetting about $I$ and $J$, in which the adoint chiral $B$ is part of the $\mathscr{N} =4$ vector multiplet.

\subsection{Higgs Branch}
Due to $\mathscr{N} = 2$ supersymmetry, the Higgs branch of the theory is a K\"{a}hler manifold. We briefly recall the construction of this space from the data defining the gauge theory. 
\subsubsection{}\label{HB}
Fix hermitian vector spaces $K \simeq \mathbb{C}^k$, $M_+ \simeq \mathbb{C}^n$, $M_- \simeq \mathbb{C}^n$. Let $(B, I, J) \in \text{End}(K) \oplus \text{Hom}(M_+, K) \oplus \text{Hom}(K, M_-)$ denote the complex scalars in the chiral multiplets. They take values in a symplectic vector space, acted on by the gauge group $U(K)$. The Higgs branch 
$$
X= X(k, n)
$$
is given by the following symplectic quotient:
\begin{equation}
X := \Big\{ (B, I, J) \in \text{End}(K) \oplus \text{Hom}(M_+, K) \oplus \text{Hom}(K, M_-) \Big| \comm{B}{B^\dagger} + II^\dagger - J^\dagger J = \zeta 1_K \Big\} \Big/U(K).
\end{equation}
The parameter $\zeta > 0$ fixing the level of the moment map is the Fayet-Iliopoulos (FI) parameter of the gauge theory. $X$ is isomorphic as a complex variety to the moduli space of degree $k$ based quasimaps to $Y = T^*Gr(n, 2n)$ in our above notation. This was explained in  \cite{nak11}. The string theory origin of the theory gives an independendent way to derive and generalize this fact \cite{ah15}.

\subsubsection{}
By a standard result (see for example chapter 3 of \cite{nakajimabook}), $X$ can be equivalently obtained as a GIT quotient, to exhibit its complex structure: 
\begin{equation} \label{xgitquot}
X = \Big\{ (B, I, J) \in \text{End}(K) \oplus \text{Hom}(M_+, K) \oplus \text{Hom}(K, M_-) \Big| \mathbb{C}[B]I(M_+) = K \Big\} \Big/GL(K). 
\end{equation}
In plain terms, the notation $\mathbb{C}[B]I(M_+) = K$ means that we require that the smallest $B$-invariant subspace of $K$ containing the image of $I$ must be $K$ itself. This is the GIT stability condition for the $GL(K)$ quotient, and is what remains of the dependence on the FI parameter $\zeta > 0$. 

Corresponding to the holomorphic isometries of $X$ are the real mass parameters of the theory, denoted $(m_\alpha^+, m_\alpha^-, \hbar)$, $\alpha = 1, \dots, n$, which come from weakly gauging the $G_F \times U(1)_\hbar$ flavor symmetry. Upon compactifying on $S^1$, both real mass and FI parameters become complexified since one can turn on holonomies around $S^1$. They are also equivariant parameters corresponding to a maximal torus $T$ of $G_F \times U(1)_\hbar$. For very small real masses, the theory flows at low energies to a supersymmetric sigma model on $X$, deformed by a real potential minimized at the $T$-fixed locus of $X$.
\subsubsection{}
For stable envelopes, it is useful to consider the subtorus $A \subset T$ acting on the variables as 
\begin{align*}
a \cdot B & = B \\
a \cdot I & = I \cdot \text{diag}(a_1^{-1}, \dots, a_n^{-1}) \\
a \cdot J & = \text{diag}(a_1, \dots, a_n) \cdot J. 
\end{align*}
Reflecting this, it is convenient to parameterize the real masses via the equivariant parameters for $A$ and another set of variables $\mu_\alpha$:
\begin{align*}
m_\alpha^+ & = a_\alpha \\
m_\alpha^- & = \hbar^{\mu_\alpha} a_\alpha. 
\end{align*}
The $a_\alpha$ parameters will eventually correspond to positions of vertex operators on the cylinder, and $\mu_\alpha$ parameters to the highest weights of Verma modules. 

\subsection{Fixed Loci}
Since we work equivariantly with respect to $T$-action on $X$, fixed points of this torus action play an essential role. They are isolated, and classified as follows. 
\subsubsection{}
For $(B, I, J)$ to represent a fixed point of $T$ action on the quotient, there must exist $g \in GL(K)$ such that 
\begin{equation}
\begin{split}
g^{-1}Bg & = \hbar^{-1} B \\
g^{-1}I & = I(\mathbf{m}^+)^{-1} \\
Jg & = \mathbf{m}^-J.
\end{split}
\end{equation}
The matrices $\mathbf{m}^\pm$ are diagonal, given by $\mathbf{m}^\pm = \text{diag}(m^\pm_1, \dots, m^\pm_n)$. 
These equations, together with the stability condition $\mathbb{C}[B]I(M_+) = K$, imply that $J = 0$, and that $K$ decomposes as a direct sum of one-dimensional eigenspaces for $g$, with eigenvalues of the form $\hbar^{i - 1}m^+_\alpha$, for $1 \leq i \leq k_\alpha$, for $n$ nonnegative integers $k_\alpha$ which sum to $k$. 

In summary, fixed points are labeled by $n$-tuples
$$\vb{k}=(k_1, \ldots, k_n)
$$
of integers $k_{\alpha}$ which sum up to $k=\sum_{\alpha=1}^n k_{\alpha}.$
A $g$ as above defines a homomorphism $T \to GL(K)$, which makes $K$ into a $T$-module, with character:
\begin{equation}\label{char}
K\eval_{\vb{k}} = \sum_{\alpha = 1}^n m_\alpha^+ \sum_{i = 1}^{k_\alpha} \hbar^{i - 1}.
\end{equation}
\subsubsection{}
We pause to explain the notation in \eqref{char}, since we will use it frequently in this paper. Since we take the quotient by $GL(K)$ in defining $X$, $K$ descends to a vector bundle over $X$, known as the tautological bundle.  Because the vector bundle $K$ carries a natural $T$-equivariant structure, we can take its equivariant K-theory class $[K] \in K_T(X)$. Since everything we do will be in K-theory, we drop the brackets and denote this again by $K$. There is a natural equivariant inclusion map of each fixed point $i_{\vb{k}}: \{ \text{pt} \} \to X$, and the pullback of $K$ under this map defines a class in $K_T(\text{pt})$, in other words, a character of the torus. It is this character that we have written in \eqref{char} -- the notation is chosen because it is literally the character of the fiber of $K$ over the fixed point $\vb{k}$.  
\subsubsection{}

To draw a parallel with Nakajima varieties, it is useful to examine in addition the $A$-fixed locus. Provided $a_\alpha$ are all distinct, the fixed locus under $A$ takes the form 
\begin{equation}
X(k, n)^A = \bigsqcup_{k_1 + \dots + k_n = k} X(k_1, 1) \times \dots \times X(k_n, 1). 
\end{equation}
Each $X(k_\alpha, 1)$ is a moduli space of degree $k_\alpha$ based quasimaps to $T^*\mathbb{P}^1=T^*Gr(1,2)$, which is a vector space. While the $A$-fixed loci are not isolated, it follows from the above discussion that they are $T$-equivariantly contractible to points. For this reason, when working in $T$-equivariant $K$-theory they are as good as fixed points. This establishes a bijection between the connected components of the $A$-fixed locus and the $T$-fixed points. This ``self-similarity'' property of $X$ with respect to taking $A$-fixed points is what gives rise to the tensor product structure in $K_T(X)$ exhibited in \eqref{rep}.

\subsubsection{}
For future purposes, it is convenient to introduce other combinatorial labelings of the fixed points. Fixed points may be enumerated by maps of finite sets $\rho: \{1, \dots, k \} \to \{1, \dots, n \}$, considered up to the action of the permutation group $S_k$. Using this action to fix an ordering $\rho(1) \geq \rho(2) \geq \dots  \geq \rho(k) \geq 1$, fixed points are labeled by partitions $\lambda = (\lambda_1 \geq \lambda_2 \geq \dots \geq \lambda_k > 0)$, of length $k$ and $\lambda_1 \leq n$, by setting $\lambda_i = \rho(i)$. 

\subsection{Supersymmetric Partition Function}
In this section, we revisit the partition function of the 3d gauge theory on ${\mathbb C}\times S^1$, in the $\Omega$-background. The partition function may be computed by first summing over contributions of maps to the pre-quotient in \eqref{xgitquot}, viewing the $G$ as a global symmetry, and then integrating over the maximal torus of $G=U(k)$. This heuristic ``Coulomb branch'' computation was done in \cite{ah15} and leads to explicit integral formulas, which are more or less known in the physics literature in the context of 3d $\mathscr{N} = 2$ indices. At first sight a less trivial claim is that the resulting partition function, for a suitable choice of contour of integration, should coincide with the K-theoretic vertex function of $X$. We first recall the heuristic computation, then compute the vertex function from the mathematical definition, and show that the two coincide as expected. 

\subsubsection{}
To state the integral formulas, it is convenient to introduce the following function, known in various places as quantum dilogarithm or reciprocal $q$-gamma function:
\begin{equation*}
\varphi_q(x) = \prod_{n = 0}^\infty (1 - q^n x) = \exp \Bigg\{ -\sum_{n = 1}^\infty \frac{x^n}{n(1 - q^n)} \Bigg\} = S^\bullet \Bigg\{ -\frac{x}{1 - q} \Bigg\}.
\end{equation*}
The final equality has been written using the plethystic exponential operation $S^\bullet$.  

The integral formulas may be most easily heuristically derived by considering the partition function on $\mathbb{C}_q \times S^1$ as a supersymmetric index with time running along $S^1$, computing the character of the Hilbert space of the theory under the maximal torus of all symmetries of the problem. Among these are the spacetime rotational symmetry with equivariant parameter $q$, and internal symmetries of the theory with equivariant parameters $m_\alpha^\pm$ and $t$; the latter corresponds to the twisted mass of the adjoint chiral multiplet containing $B$. It is related to the parameter $\hbar$ in the previous section by $t = q/\hbar$. To preserve supersymmetry, a certain $R$-symmetry twist is necessary as well; equivalently we work in a partially twisted version of the theory.  

Since such an index is independent of the 3d gauge coupling, it may be reliably computed in the limit that the theory becomes free, where the vector multiplet is frozen out and the gauge symmetry may be abelianized and considered as a global symmetry, with equivariant parameters given by the Coulomb moduli. The fact that that the vector multiplet is dynamical means that we must impose Gauss' law and project to invariants, leading to integration over the Coulomb moduli. As this computation is simply a free-field character, all contributions may be expressed as $S^\bullet$ of single-particle characters, leading to simple expressions in terms of $\varphi_q$-functions. 

The vector multiplet and adjoint chiral multiplet containing $B$ contribute 
\begin{equation}
\prod_{1 \leq i \neq j \leq k} \frac{\varphi_q(x_i/x_j)}{\varphi_q(tx_i/x_j)}.
\end{equation}
The fundamental and antifundamental chiral multiplets containing $I$ and $J$ from section \ref{HB} contribute  the numerator and the denominator in:
\begin{equation}
\prod_{\alpha = 1}^n \prod_{i = 1}^k \frac{\varphi_q(qx_i/m_\alpha^-)}{\varphi_q(x_i/m_\alpha^+)}. 
\end{equation}
Combining these pieces, and recalling the contribution of the FI term, the partition function is given by the contour integral 
\begin{equation}
Z_{3d}[\gamma] = \int_\gamma [dx] \prod_{i = 1}^k x_i^{-\frac{\log(z)}{\log (q)}} \prod_{1 \leq i \neq j \leq k} \frac{\varphi_q(x_i/x_j)}{\varphi_q(tx_i/x_j)} \prod_{\alpha = 1}^n \prod_{i = 1}^k \frac{\varphi_q(qx_i/m_\alpha^-)}{\varphi_q(x_i/m_\alpha^+)}. 
\end{equation}
$[dx]$ denotes the Haar measure on the torus $U(1)^k$, regarded as a middle dimensional cycle in $(\mathbb{C}^\times)^k$, and we introduced $z = e^{-\zeta}$. 

\subsubsection{}
To fully specify the partition function one should make a choice of integration contour $\gamma$. These are closely related to boundary conditions at infinity of $\mathbb{C}$ in ${\mathbb C}$ in ${\mathbb C}\times S^1$. A better perspective is to keep the contour of integration fixed and generate boundary conditions by coupling to a two dimensional $(0, 2)$ theory on the boundary. At the practical level, this corresponds to insertion of additional theta functions in the integral, from the elliptic genus of the boundary theory. The elliptic genera of boundary theories determine classes in $\text{Ell}_{T}({X})$, the equivariant elliptic cohomology of $X$. 

One of our key results is a construction of very special classes in $\text{Ell}_{T}({ X})$, the elliptic stable envelopes, which we discuss in section \ref{ellipticstabandpoles}. At the moment, we will focus on the behavior in the bulk.
\subsubsection{}

For $|z| < 1$ ($\zeta > 0$), the integral may be evaluated by residues, closing the contour to pick up the sequences of poles such that the corresponding series of terms weighted by powers of $z$ is convergent. For $|q| < 1$, these correspond to sequences of poles accumulating at $|x| \to \infty$. The boundary condition at infinity of ${\mathbb C}$ chooses which poles contribute.

Geometrically, the poles correspond to fixed points in the moduli space of quasimaps $\mathbb{P}^1 \to X$ (of all degrees), and the residues are the localization contributions. 
The proof that the integral formula captures enumerative geometry of quasimaps, and that the gauge theory partition function coincides with the vertex function of $X$, will be given in the next section. 
 
\subsubsection{}
Integral formulas also exist in the presence of insertions, that is, for computations of expectation values of operators in the gauge theory. The Wilson line operators $\mathscr{O}_R$ introduced below in \eqref{Wl} become the $U(k)$ characters evaluated on $x = \text{diag}(x_1, \dots, x_k)$, which are well-known to generate the ring of symmetric Laurent polynomials in $x_1, \dots, x_k$. 

Any rational function $f(x)$ may be inserted into the integral, to obtain an expression known as a vertex with descendent. Geometrically, the integration variables are identified with Chern roots of the tautological bundle $K \to X$, so the descendent insertions correspond to certain lifts of classes in $K_T(X)$ (more precisely, a certain localized K-theory if one allows for denominators) to the ambient quotient stack in which $X$ is an open subset, see \cite{ao17}. 

\subsection{Integral Formula from $(2 + 1)$-dimensional Instanton Counting}
In the next few sections we recall the derivation of the integral formula given above from instanton counting in the gauged linear sigma model with Higgs branch $X$. That such an integral formula exists for the instanton sum is well-known in the physics literature. In what follows, we give a rigorous proof of this result by directly performing the localization computation and matching with residues. See also the recent work \cite{nekrasovdedushenko2}, \cite{crew2023} for a related discussion of the kind of partition functions considered here.

While this computation is necessarily somewhat technical, the key takeaway is that many of the features of Nakajima varieties responsible for the rich structure present in their enumerative theory of quasimaps in fact persist in the present setting. This leads to integral formulas with a more symmetric structure than would be found in a generic $\mathscr{N} = 2$ theory. The key technical concept seems to be the notion of polarization, in the terminology of \cite{okounkov15, okounkov20, okounkov21}.

\subsubsection{}
As recalled above, at low energies, and when the FI parameter $\zeta$ is large, this theory reduces to a nonlinear sigma model with target $X$. 

The partition function of the nonlinear sigma model to $X$ in the 3d $\Omega$-background $\mathbb{C}_q \times S^1$ can be evaluated by standard techniques of supersymmetric localization, and reduces formally to supersymmetric quantum mechanics on $S^1$ with target given by the moduli space of instantons in the sigma model, that is, holomorphic maps from $\mathbb{C} \to X$ (or equivalently, the space of based rational maps $\mathbb{P}^1 \to X$, where a choice of basepoint for the maps is made at $\infty \in \mathbb{P}^1$ reflecting the need to fix asymptotic behavior of fields on $\mathbb{C}$). The supersymmetric quantum mechanics in question computes the equivariant index of the Dirac operator on the instanton moduli space, coupled to the (obstruction) bundle of fermion zero modes. 

\subsubsection{}
The moduli space of holomorphic maps $\mathbb{P}^1 \to X$ is not satisfactory for computation because it is noncompact, due to the fact that holomorphic maps may degenerate in families. This issue of compactifying the moduli space is closely related to the need to introduce a UV completion of the nonlinear sigma model to $X$; in this case, the UV completion is simply the 3d $\mathscr{N} = 2$ gauge theory we started with, and the compactification corresponding to it is known as quasimaps (see \cite{okounkov15} for an introduction to basic facts about quasimaps, and \cite{lns2000} for more on the physics of this). In physical language, the quasimaps are vortices in the gauge theory, and the singularities of a quasimap correspond to a vortex shrinking to its minimum allowed size. 

Computing the partition function of the theory is then reduced to the well-defined mathematical problem of K-theoretic equivariant integration over the moduli space of quasimaps $QM_p(X)$---specifically, we consider the moduli space of quasimaps from $\mathbb{P}^1$ to $X$ approaching a chosen point $p \in X^T$ at $\infty \in \mathbb{P}^1$, whence the notation. This moduli space splits into a countable union of finite-dimensional connected components labeled by an integer $d \geq 0$ referred to as degree. The degree $d$ is also the vortex charge. We denote by $QM_p(X; d)$ the connected component formed by degree $d$ quasimaps. 

The torus $\mathbb{C}^\times_q \times T$ acts on $QM_p(X)$, where the first factor acts by automorphisms of the base curve $\mathbb{P}^1$ and the action of $T$ is induced by the action on $X$. We consider enumerative geometry computations in equivariant K-theory, $K_{\mathbb{C}^\times_q \times T}(QM_p(X))$. Equivariance with respect to $T$ corresponds to turning on (real) mass parameters in the gauge theory.
\subsubsection{}
For each fixed point $p \in X^T$, the partition function of the model is 
\begin{equation}
Z_p = \sum_{d \geq 0} z^d \chi(QM_p(X; d), \widehat{\mathscr{O}}_{\text{vir}})    
\end{equation}
where the object $\widehat{\mathscr{O}}_{\text{vir}}$ is an equivariant K-theory class on $QM_p(X)$ known as the symmetrized virtual structure sheaf, which will be defined below, and $\chi$ denotes equivariant Euler characteristic. The parameter $z = e^{-\zeta}$ is the instanton counting/FI parameter. $Z_p$ is a particular component of the K-theoretic vertex function of $X$ \cite{okounkov15}, expressed in the fixed point basis. 

Specifically, the quantity above is the instanton part of the partition function---the integral formula also includes certain one-loop prefactors, which we specify below. 

In the subsequent sections we spell out the definition of the symmetrized virtual structure sheaf, and the localization formula for $Z_p$. 

\subsection{Quasimaps and $\widehat{\mathscr{O}}_{\mathrm{vir}}$}
In this section we review basics on quasimaps. This material is well-known and reviewed in \cite{okounkov15}, but we recall it to establish notations and make some small modifications to the standard story necessary since $X$ is not a Nakajima quiver variety. 

\subsubsection{}
In the previous section, we recalled that $X$ could be described as a GIT quotient of a space of matrices $(B, I, J)$ acting on vector spaces $K$ and $M_{\pm}$, where the quotient was taken by $GL(K)$. By definition, a quasimap from $\mathbb{P}^1$ to $X$ is a vector bundle $\mathscr{K}$ on $\mathbb{P}^1$ of rank $k$, together with two trivial vector bundles $\mathscr{M}_\pm$ of rank $n$ and a section 
\begin{equation}
f = (\mathscr{B}(z), \mathscr{I}(z), \mathscr{J}(z)) \in H^0( \mathscr{K} \otimes \mathscr{K}^* \oplus \mathscr{K} \otimes \mathscr{M}^*_+ \oplus \mathscr{M}_- \otimes \mathscr{K}^*).     
\end{equation}
The value $f(p)$ of a quasimap at a point $p \in \mathbb{P}^1$ is well-defined as a $GL(K)$ orbit in the prequotient, or equivalently as a point in the quotient stack in which $X$ embeds naturally. A quasimap $f$ is called stable if $f$ takes values in the stable locus at all but finitely many points of $\mathbb{P}^1$---such points are referred to as the singularities of $f$. The degree of a quasimap is by definition the degree of the bundle $\mathscr{K}$, $\deg f = \deg \mathscr{K}$. 

The moduli space $QM_p(X; d)$ consists of those which satisfy $f(\infty) = p$ and $\deg \mathscr{K} = d$. At a generic point of the moduli space, $f$ determines an honest holomorphic map from $\mathbb{P}^1 \to X$, $\mathscr{K}$ is pulled back from the tautological bundle $K$ on $X$, and $d$ is the degree of $f$ as a holomorphic map. 
\subsubsection{}
Quasimaps come with a natural obstruction theory, which arises naturally from gauge theory by studying the zero modes of all fields in a given instanton background. The bundles $\mathscr{K}$ and $\mathscr{M}_\pm$ define universal bundles on $QM_p(X) \times \mathbb{P}^1$ of the same name. The virtual tangent space of $QM_p(X)$ is the K-theory class (cohomology is taken along $\mathbb{P}^1$): 
\begin{equation}
T_{\text{vir}} = H^\bullet(\mathscr{K} \otimes \mathscr{M}^*_+ \oplus \mathscr{M}_- \otimes \mathscr{K}^*) + (t - 1)\text{Ext}^\bullet(\mathscr{K}, \mathscr{K}) - \text{ev}_\infty^*(TX \big|_p) \in K_{\mathbb{C}^\times_q \times T}(QM_p(X)).     
\end{equation}
The equivariant parameter $t$ is the mass of the field $B$,  related to the parameter $\hbar$ used in section \ref{quivergauge} by $t=q/\hbar$. The bullet is used to denote the alternating sum of $0$ and $1$ cohomology groups, as in \cite{okounkov15}. Each term in the virtual tangent space arises as follows: $H^0$ and $t\text{Ext}^0$ arise from deformations of the maps $(\mathscr{B}, \mathscr{I}, \mathscr{J})$, $-H^1$ and $-t\text{Ext}^1$ are the fermion zero modes, $\text{Ext}^1$ are the deformations of the bundle related to the vector multiplet in gauge theory, and $-\text{Ext}^0$ are the gauge transformations. $\text{ev}_\infty^*$ denotes pullback under evaluation at infinity, which in this case just pulls back fibers of bundles at the point $p$. 
\subsubsection{}

This obstruction theory defines a virtual structure sheaf $\mathscr{O}_{\text{vir}}$. A key technical point in this paper is that $X$ admits a polarization, which is in some sense a square root of its tangent bundle. Define the polarization 
\begin{equation} \label{Tsqrt}
\mathscr{T}^{1/2} = \mathscr{K} \otimes \mathscr{M}^*_+.
\end{equation}
The symmetrized virtual structure sheaf is defined as:
\begin{equation}
\widehat{\mathscr{O}}_{\text{vir}} := \mathscr{O}_{\text{vir}} \otimes \Bigg( \mathscr{K}_{\text{vir}} \otimes \frac{\det \mathscr{T}^{1/2} \eval_\infty}{\det \mathscr{T}^{1/2} \eval_0} \Bigg)^{1/2},     
\end{equation}
where $\mathscr{K}_{\text{vir}} = \det T^*_{\text{vir}}$ denotes virtual canonical bundle.
It is easy to check, as in section 6.1 of \cite{okounkov15}, that the square root exists after allowing for square roots of the equivariant parameters $m_\alpha^-/m_\alpha^+$. At the practical level, the role of the polarization in this formula is simply to avoid having to take square roots of $q$ when performing localization computations. 

There is also a canonical identification 
\begin{equation}
\text{ev}^*_\infty(TX \big|_p) = (\mathscr{K} \otimes \mathscr{M}_+^* + \mathscr{M}_- \otimes \mathscr{K}^* + (t - 1)\mathscr{K} \otimes \mathscr{K}^*)\eval_\infty. 
\end{equation}

\subsection{Localization}
Now we turn to computation of $Z_p$ by localization and complete the identification of $Z_p$ with the contour integral above, for particular choice of contour $\gamma$. It will be convenient to view $p \in X^T$ as an $n$-tuple of integers $\vb{k} = (k_1, \dots, k_n)$ such that $k = \sum_\alpha k_\alpha$. 
\subsubsection{}
The first step of a localization computation is to analyze the torus fixed points on $QM_p(X; d)$. A $\mathbb{C}^\times_q \times T$-invariant quasimap $f$, which is nonsingular at infinity, must have all its singularities at $0$ and evaluate to $p$ on $\mathbb{P}^1 \setminus \{ 0 \}$. One can use the freedom of gauge transformations to identify $\mathscr{K}$ with a sum of line bundles: 
\begin{equation}
 \mathscr{K} = \bigoplus_{\alpha = 1}^n \bigoplus_{i = 1}^{k_\alpha} \mathscr{O}(\lambda^{(\alpha)}_i).    
\end{equation}
The quantities $\lambda^{(\alpha)}_i$ are, at this stage, integers which sum to $d$ but are otherwise arbitrary. The bundles $\mathscr{M}_\pm$ are trivial topologically but have nontrivial $T$-equivariant structures, and the bundle maps $(\mathscr{B}(z), \mathscr{I}(z), \mathscr{J}(z))$ must be $\mathbb{C}^\times_q \times T$-equivariant. 

With the above splitting of $\mathscr{K}$ fixed, all bundles are naturally trivial on $\mathbb{P}^1 \setminus \{ \infty \}$. The fixed behavior of sections away from $0$ and equivariance requirement is enough to determine them uniquely, and they take the following form. It is convenient to write $\mathscr{K} = \oplus_\alpha \mathscr{K}_\alpha$, where $\mathscr{K}_\alpha = \oplus_i \mathscr{O}(\lambda^{(\alpha)}_i)$. Then, for $z \in \mathbb{P}^1 \setminus \{ \infty \}$, sections are of the form (the index $i = 1, \dots, k_\alpha$): 
\begin{align}
\mathscr{B}(z) \eval_{\mathscr{K}_\alpha} & = \begin{pmatrix} 0 & z^{\lambda^{(\alpha)}_1 - \lambda^{(\alpha)}_2} & \dots \\
 & 0 & z^{\lambda^{(\alpha)}_2 - \lambda^{(\alpha)}_3} & \dots \\
  &  & \ddots   & \vdots \\ & & \dots & 0  \end{pmatrix} \\
\mathscr{I}\indices{^i_\alpha}(z) & = z^{\lambda^{(\alpha)}_i} \delta^{i, k_\alpha}  \\
\mathscr{J}(z) & = 0. 
\end{align}
The condition that these sections are holomorphic is the same as $\lambda^{(\alpha)}_1 \geq \lambda^{(\alpha)}_2 \geq \dots \geq \lambda^{(\alpha)}_{k_{\alpha}} \geq 0$, so that $\lambda^{(\alpha)}$ comprise an $n$-tuple of partitions, with the constraint $\ell(\lambda^{(\alpha)}) \leq k_\alpha$ where $\ell$ denotes length of a partition. The degree of the quasimap is given by 
\begin{equation}
d = \deg f = \deg \mathscr{K} = \sum_{\alpha = 1}^n \sum_{1 \leq i \leq k_\alpha} \lambda^{(\alpha)}_i    
\end{equation}
that is, the total number of boxes. The conclusion is that the $\mathbb{C}^\times_q \times T$-fixed points on $QM_p(X; d)$ are in one to one correspondence with $n$-tuples of partitions $\lambda^{(\alpha)}$, with $d$ total boxes and $\ell(\lambda^{(\alpha)}) \leq k_\alpha$, where the tuple $(k_1, \dots, k_n)$ is determined by $p \in X^T$. We adopt the notation $\underline{\lambda}$ for an $n$-tuple of partitions, and $d = |\underline{\lambda}|$ for the number of boxes. 
\subsubsection{}
The next step is to determine the localization weight associated to each fixed point. The quantity we wish to compute is the Euler character of $\widehat{\mathscr{O}}_{\text{vir}}$, which up to less important factors is the Euler character of $\mathscr{O}_{\text{vir}} \otimes \mathscr{K}^{1/2}_{\text{vir}}$. The contribution of a fixed point $\underline{\lambda}$ to $Z_p$ is 
\begin{equation} \label{weight}
z^{|\underline{\lambda}|} \widehat{a} \Bigg(T_{\text{vir}} \eval_{\underline{\lambda}} \Bigg) \times q^{\#}
\end{equation}
where the function $\widehat{a}(x)$ is 
\begin{equation}
\widehat{a}(x) = \frac{1}{x^{1/2} - x^{-1/2}}
\end{equation}
and is extended to K-theory as a genus, that is, for a virtual bundle $\mathscr{E} - \mathscr{F}$
\begin{equation}
\widehat{a}(\mathscr{E} - \mathscr{F}) = \prod_{x \in \text{Chern roots of $\mathscr{E}$}} \frac{1}{x^{1/2} - x^{-1/2}} \prod_{y \in \text{Chern roots of $\mathscr{F}$}} (y^{1/2} - y^{-1/2}). 
\end{equation}
The power of $q$ in \eqref{weight} ensures that the total quantity is defined without having to take square roots, and is the contribution of $\det \mathscr{T}^{1/2}$ to $\widehat{\mathscr{O}}_{\text{vir}}$. 

The class $T_{\text{vir}}$ restricted to a fixed point is an element of $K_{\mathbb{C}^\times_q \times T}(\text{pt})$, namely a character of the torus, and the only remaining problem is to determine which one. This is a standard exercise in equivariant Grothendieck-Riemann-Roch theorem, applied in this instance to the natural projection $\pi: QM_p(X; d) \times \mathbb{P}^1 \to QM_p(X; d)$: 
\begin{align*}
R\pi_* & \Big( \mathscr{K} \otimes \mathscr{M}_+^* \oplus \mathscr{M}_- \otimes \mathscr{K}^* \oplus (t - 1)\mathscr{K} \otimes \mathscr{K}^* \Big) \\
& = \frac{(\mathscr{K} \otimes \mathscr{M}_+^* + \mathscr{M}_- \otimes \mathscr{K}^* + (t - 1)\mathscr{K} \otimes \mathscr{K}^*) \eval_0}{1 - q} + \frac{(\mathscr{K} \otimes \mathscr{M}_+^* + \mathscr{M}_- \otimes \mathscr{K}^* + (t - 1)\mathscr{K} \otimes \mathscr{K}^*) \eval_\infty}{1 - q^{-1}}. 
\end{align*}
Then it is clear that, upon subtracting $ev_\infty^*(TX \big|_p)$, 
\begin{equation} \label{Tvirloc}
T_{\text{vir}} = \frac{(\mathscr{K} \otimes \mathscr{M}_+^* + \mathscr{M}_- \otimes \mathscr{K}^* + (t - 1)\mathscr{K} \otimes \mathscr{K}^*) \eval_0}{1 - q} - \frac{(\mathscr{K} \otimes \mathscr{M}_+^* + \mathscr{M}_- \otimes \mathscr{K}^* + (t - 1)\mathscr{K} \otimes \mathscr{K}^*) \eval_\infty}{1 - q}. 
\end{equation}
Upon restriction to the fixed point $\underline{\lambda}$, this can be made particularly explicit. As a consequence of the analysis of the fixed loci, we have (we omit the explicit restriction to $\underline{\lambda}$ to lighten notation, but it is understood throughout): 
\begin{align}
\mathscr{K}\eval_0 & = \sum_{\alpha = 1}^n m_\alpha^+ \sum_{i = 1}^{k_\alpha} t^{i - k_\alpha} q^{-\lambda^{(\alpha)}_i} \\
\mathscr{K}\eval_\infty & = \sum_{\alpha = 1}^n m_\alpha^+ \sum_{i = 1}^{k_\alpha} t^{i - k_\alpha} \\
\mathscr{M}_\pm \eval_{0, \infty} & = \sum_{\alpha = 1}^n m^\pm_\alpha. 
\end{align}
\subsubsection{}

As a consequence of the compactness of $\mathbb{P}^1$, upon restriction to $\underline{\lambda}$ \eqref{Tvirloc} is a Laurent polynomial in $q$. Applying $\widehat{a}$ to it gives the contribution of $\underline{\lambda}$ to the localization formula, and the result may be put into the following standard form. We have: 
\begin{multline}
\widehat{a} (T_{\text{vir}} - \text{ev}^*_\infty(TX \big|_p)) = q^{-nd/2} \prod_{\alpha = 1}^n (-m_\alpha^-/m_\alpha^+)^{d/2} \times \\ \frac{\varphi_q(q \mathscr{K} \otimes \mathscr{M}_-^* \eval_0) \varphi_q(\mathscr{K} \otimes \mathscr{K}^* \eval_0)}{\varphi_q(\mathscr{K} \otimes \mathscr{M}_+^* \eval_0) \varphi_q(t\mathscr{K} \otimes \mathscr{K}^* \eval_0)} \times \frac{\varphi_q( \mathscr{K} \otimes \mathscr{M}_+^* \eval_\infty) \varphi_q(t\mathscr{K} \otimes \mathscr{K}^* \eval_\infty)}{\varphi_q(q \mathscr{K} \otimes \mathscr{M}_-^* \eval_\infty) \varphi_q(\mathscr{K} \otimes \mathscr{K}^* \eval_\infty)} 
\end{multline}
where we have extended $\varphi_q(x)$ to K-theory as a genus in a similar fashion. The power of $q$ exactly cancels the power of $q$ from $\det \mathscr{T}^{1/2}$ in $\widehat{\mathscr{O}}_{\text{vir}}$, and ratios of masses may be absorbed into a redefinition of $z$ (this is a one-loop renormalization of the FI parameter).

\subsubsection{}

Using the above formulas, it is then easy to express the contributions of fixed points in terms of $\varphi_q$-functions, with the result that the instanton partition function of the model is given by 
\begin{equation} \label{normalization}
 Z_p = \prod_{\alpha, \beta = 1}^n \prod_{i = 1}^{k_\alpha} \frac{\varphi_q(\frac{m_\alpha^+}{m_\beta^+} t^{i - k_\alpha})}{\varphi_q(\frac{m_\alpha^+}{m_\beta^-} q t^{i - k_\alpha})}\prod_{\alpha, \beta = 1}^n \prod_{i = 1}^{k_\alpha} \prod_{j = 1}^{k_\beta} \frac{\varphi_q( \frac{m_\alpha^+}{m_\beta^+} t^{1 + i - j - k_\alpha + k_\beta})}{\varphi_q( \frac{m_\alpha^+}{m_\beta^+} t^{i - j - k_\alpha + k_\beta})} \times Z_{\vb{k}}
\end{equation}
where
\begin{equation}
Z_{\vb{k}} = \sum_{\substack{\underline{\lambda} \\ \ell(\lambda^{(\alpha)}) \leq k_\alpha}} z^{|\underline{\lambda}|} \prod_{\alpha, \beta = 1}^n \prod_{i = 1}^{k_\alpha} \frac{\varphi_q(\frac{m_\alpha^+}{m_\beta^-} t^{i - k_\alpha} q^{1 - \lambda^{(\alpha)}_i})}{\varphi_q(\frac{m_\alpha^+}{m_\beta^+} t^{i - k_\alpha}q^{-\lambda^{(\alpha)}_i})} \prod_{\alpha, \beta = 1}^n \prod_{i = 1}^{k_\alpha} \prod_{j = 1}^{k_\beta} \frac{\varphi_q( \frac{m_\alpha^+}{m_\beta^+} t^{i - j - k_\alpha + k_\beta}q^{\lambda^{(\beta)}_j - \lambda^{(\alpha)}_i})}{\varphi_q( \frac{m_\alpha^+}{m_\beta^+} t^{1 + i - j - k_\alpha + k_\beta}q^{\lambda^{(\beta)}_j - \lambda^{(\alpha)}_i})}.
\end{equation}

As written, both of these expressions are ill-defined because various $\varphi_q$-functions are zero, however the zeroes exactly cancel between all numerators and denominators. That is to say, it is understood that each term in $Z_{\vb{k}}$ is evaluated using the $q$-difference equation 
\begin{equation*}
\varphi_q(q^{-1}x) = (1 - q^{-1}x) \varphi_q(x)
\end{equation*}
and any $\varphi_q(1)$ that appears is canceled against the $\varphi_q(1)$ in the prefactor of \eqref{normalization}. 
\subsubsection{}
When the result is written in this form, it is clear that up to overall normaliations it is equivalent to the prescription of evaluation the integral 
\begin{equation} \label{partitionfn}
Z = \int_\gamma \prod_{j = 1}^k \frac{dx_j}{2\pi i x_j} \prod_{j = 1}^k x_j^{-\frac{\log(z)}{\log(q)}} \prod_{i \neq j = 1}^k \frac{\varphi_q(x_i/x_j)}{\varphi_q(tx_i/x_j)} \prod_{\alpha = 1}^n \prod_{i = 1}^k \frac{\varphi_q(qx_i/m_\alpha^-)}{\varphi_q(x_i/m_\alpha^+)}
\end{equation}
by residues, where the contour $\gamma$ is chosen to enclose the following sequence of poles of the integrand. Split the $k$ variables $x_i$ into $n$ groups of size $k_\alpha$, and label the corresponding variables in the group by $x^{(\alpha)}_i$, $i = 1, \dots, k_\alpha$. Since the integrand is a symmetric function of the $x_i$, such a splitting is purely a bookkeeping device. The poles are given by 
\begin{equation}
x^{(\alpha)}_i = m_\alpha^+ q^{-\lambda^{(\alpha)}_i}t^{i - k_\alpha}    
\end{equation}
and $\lambda^{(\alpha)}_1 \geq \dots \geq \lambda^{(\alpha)}_{k_\alpha} \geq 0$. In this way the pole structure of the integrand reproduces the structure of fixed points, and the above analysis of equivariant localization contributions establishes that the integral does indeed give rise to a K-theoretic count of quasimaps to $X$ when evaluated by residues, proving the claim made above. The normalization of $Z$ provided by the integral is the one natural from the perspective of both $q$-difference equations and quantum field theory. In quantum field theory langauge, it also includes the tree level and one-loop contributions to the $\Omega$-background partition function, while the quasimap enumeration corresponds to the instanton sum only. 

\subsubsection{}
The contour $\gamma$ is related to the boundary conditions of the theory at infinity---the above choice is equivalent to a 0-brane boundary condition in the sigma model to $X$, where the 0-brane is placed at the fixed point $p \in X^T$ corresponding to the tuple $k_\alpha$. For the simple 0-brane boundary conditions, the identification of a choice of boundary condition with a choice of integration contour is convenient, but later we will need other boundary conditions described by the elliptic stable envelopes. These are best described by coupling the theory to boundary degrees of freedom; this will be described in greater detail in a future section. 

\section{Stable Envelopes, Defects, and Boundary Conditions} \label{stabs}

In the present section we explain the construction of a certain canonical basis in the twisted chiral ring of the 3d $\mathscr{N} = 2$ theory under consideration. First we recall some basic facts on twisted chiral rings and their relation to the geometry of the Higgs branch. 

\subsection{Twisted Chiral Ring}\label{TCR}
A three-dimensional $\mathscr{N} = 2$ theory compactified on $S^1$ may be viewed as a two-dimensional $\mathscr{N} = (2, 2)$ theory with an infinite number of fields. The latter can be subjected to the A-type topological twist, and one can study the twisted chiral ring of the model, that is, the cohomology of the A-model supercharge $Q_A$ in the space of local operators in two dimensions. Importantly, this same supercharge gives rise to the localization onto quasimaps, leading to integral formulas reviewed above.  

Any 3d $\mathscr{N} = 2$ gauge theory with gauge group $U(k)$ has a real adjoint scalar $\varphi$ in its vector multiplet. Consider our theory on the geometry $D \times S^1$, where $D$ has the topology of a disk. If $A_t$ denotes the component of the gauge field along the $S^1$ direction, and $R$ is a representation of $U(k)$, the supersymmetric Wilson line operators 
\begin{equation}\label{Wl}
\mathscr{O}_R(x) = \tr_{R} \mathcal{P} \exp \Bigg( \oint_{S^1} dt(A_t(x, t) + i \varphi(x, t)) \Bigg)
\end{equation}
generate the twisted chiral ring of the model viewed as a 2d $\mathscr{N} = (2, 2)$ theory. The twisted chiral ring may be identified with the equivariant quantum $K$-theory of the Higgs branch, which is in turn identified as a vector space with the space of supersymmetric vacua of the theory. After flowing to the nonlinear sigma model on $X$, as classes in $K_T(X)$, these operators are identified with the classes of the bundles associated to the tautological bundle $K \to X$ via the representation $R$. 

\subsubsection{}
Quite generally, codimension two defects of the three dimensional gauge theory at hand, supported on the $S^1$, define particular objects in the chiral ring of the effective two dimensional theory on ${\mathbb C}$. Suitably chosen defects will  define a certain very canonical basis of classes in $K_T(X)$ (here we think of the K-theory as a vector space, ignoring the ring structure). The stable basis, which we construct in this section, generalizes the stable basis known to exist for Nakajima varieties. 
 
We will also show that, inserting the K-theoretic stable envelopes into the integral formulas as descendent insertions, produces the $q$-conformal blocks of $\mathscr{U}_\hbar(\widehat{\mathfrak{sl}_2})$, solutions to the quantum Knizhnik-Zamolodchikov equations with vertex operators corresponding to Verma modules of $\mathscr{U}_\hbar(\mathfrak{sl}_2)$. 

\subsection{Defects from Orbifolds}
The defects we are interested in studying may be defined as singular boundary conditions on the bulk fields of the theory associated to a codimension two locus in spacetime. Specifically, a submanifold of codimension two can be linked by a loop, so the gauge fields are required to have a fixed conjugacy class of holonomy around arbitrarily small loops encircling the worldvolume of the defect. Similar singular behavior is prescribed for the matter fields.
Alternatively, we may also define the defect by coupling the bulk theory to a sigma model living on the defect. Integrating out the fields on the worldvolume of the defect leads to the singular behavior of fields in the bulk. These conceptual definitions of the supersymmetric defects are often not useful for performing exact computations.

An exception, explained by Nekrasov \cite{nek18}, is when the desired singular behavior may be engineered by an orbifold procedure, where the orbifold is performed with respect to a cyclic group $\mathbb{Z}_p$ acting in the obvious way on the $\mathbb{C}$ factor of the spacetime $\mathbb{C} \times S^1$. 

\subsubsection{}
The basic idea \cite{nek18} is that, because $(\mathbb{C} \times S^1)/\mathbb{Z}_p \simeq \mathbb{C} \times S^1$, the gauge theory on the orbifold will be the same as the gauge theory ``upstairs'' , with a defect inserted at the fixed set of the orbifold group action. We only need a choice of an embedding of the orbifold group $\mathbb{Z}_p$ into the gauge and flavor groups. The $\mathbb{Z}_p$-invariance ``upstairs'' is reinterpreted downstairs as the singular boundary condition on the fields, generating the defect at $\{0\} \times S^1$. 

For the theories under consideration, take $p = n(k + 1)$. It is convenient to change the notation and replace the variables $x_i, q, m_\alpha^\pm$ in the upstairs theory by $\tilde{x}_i, \tilde{q}, \tilde{m}_\alpha^\pm$. Introduce also the characters of various spaces under the $T$-action (we abuse notation and identify the vector space with the character---this makes the calculations which follow more economical, and is of course well-motivated by K-theory as explained previously):
\begin{equation}
\begin{split}
\widetilde{K} & = \sum_{i = 1}^k \tilde{x}_i \\
\widetilde{M}^\pm & = \sum_{\alpha = 1}^n \tilde{m}_\alpha^\pm. 
\end{split}
\end{equation}
The action of $\Gamma=\mathbb{Z}_{n(k + 1)}$ on the spacetime induces the action $\tilde{q} \mapsto \omega \tilde{q}$, for $\omega=e^{2\pi i \over n(k+1)}$ a generator of $\Gamma$. To fully define the orbifold action, the spaces $\widetilde{K}$, $\widetilde{M}^\pm$ are made into $\Gamma$ modules: 
\begin{equation}
\begin{split}
\widetilde{K} & = \bigoplus_{j = 1}^{n(k + 1)} \widetilde{K}_j \otimes R_j \\
\widetilde{M}^\pm & = \bigoplus_{j = 1}^{n(k + 1)} \widetilde{M}^\pm_j \otimes R_j
\end{split}
\end{equation}
where $R_j$ is the representation $\omega \mapsto \omega^j$, and the coefficients are multiplicity spaces. Introduce the notation $[N]$ for the finite set $\{1, \dots, N \}$. 
\subsubsection{}
To relate the two decompositions (as $T$ and $\Gamma$ modules), introduces the coloring functions $c_K: [k] \to [n(k + 1)]$, $c_\pm: [n] \to [n(k + 1)]$, defined by the relations
\begin{equation}
\begin{split}
\widetilde{K}_j & = \sum_{i \in c^{-1}_K(j)} \tilde{x}_i \\
\widetilde{M}^\pm_j & = \sum_{\alpha \in c^{-1}_\pm(j)} \tilde{m}^\pm_\alpha.
\end{split}
\end{equation}
Given a fixed point labeled by a partition $\lambda = (\lambda_1 \geq \dots \geq \lambda_k > 0)$, the coloring functions are defined as follows:
\begin{equation}
\begin{split}
c_-(\alpha) & = \alpha + k(\alpha - 1) \\
c_+(\alpha) & = \alpha(k + 1) \\
c_K(i) & = (k + 1)\lambda_i - \#\{ j < i | \lambda_j = \lambda_i \}. 
\end{split}
\end{equation}
It is clear from definitions that $i < j$ is equivalent to $c_K(i) > c_K(j)$. This completes the specification of the orbifold action defining the defect; it is clear combinatorially that there is a single defect for each connected component of $X^T$. 

\subsubsection{}
In English, what is happening here is the following. The structure of the spaces $\widetilde{K}$, $\widetilde{M}_\pm$ as $\Gamma$-modules determines how the fields of the theory will transform under the orbifold group, which translates ``downstairs'' to a certain prescribed singular behavior. As we will see shortly via explicit computation, this singular behavior is determined by the relative values of the $c$-functions in $[n(k + 1)]$, regarded as a discrete approximation to the real line. 

The $c$-functions are assigned to mass parameters, as is clear above, so correspondingly $[n(k + 1)]$ approximates the real mass line. For this reason, orbifolding by $\Gamma = \mathbb{Z}_p$ for any $p$ will do, so long as $p \gg 0$; we have chosen the minimal $p$ for our purposes. What we are lead to is a class of defects which depend on the relative positions of masses along the real mass line---as it will turn out, these are none other than K-theoretic stable envelopes of \cite{okounkov15} (as a comment for the experts: stable envelopes at zero slope). 

\subsubsection{}
The partition function in the presence of the defect is simply the path integral over the $\Gamma$-invariant fields. It is straightforward to present its integral representation: the integrand is (ignoring for the moment the contribution of the FI term, and it is understood throughout that the diagonal contributions in $KK^*$ are deleted)
\begin{equation}
S^\bullet \Bigg\{ \Bigg( \frac{(\hbar^{-1} \tilde{q} - 1)}{1 - \tilde{q}} \widetilde{K} \widetilde{K}^* + \frac{\widetilde{K}}{1 - \tilde{q}}(\widetilde{M}^{+ *} - \tilde{q} \widetilde{M}^{-*}) \Bigg)^{\Gamma} \Bigg\}.
\end{equation}
This must be expressed in terms of the $\Gamma$-invariant variables 
\begin{equation*}
\begin{split}
q & := \tilde{q}^{n(k + 1)} \\
m_\alpha^\pm & := \tilde{m}_\alpha^\pm \tilde{q}^{-c_\pm(\alpha)} \\
x_i & := \tilde{x}_i \tilde{q}^{-c_K(i)}, 
\end{split}
\end{equation*}
making use of the elementary identity
\begin{equation*}
\frac{1}{1 - \tilde{q}} = \frac{1}{1 - q} \sum_{\ell = 0}^{n(k + 1) - 1} \tilde{q}^\ell.
\end{equation*}

\subsubsection{}
For example, the contribution of the vector multiplet is 
\begin{equation}
\begin{split}
\Bigg(\frac{1}{1 - \tilde{q}}\widetilde{K} \widetilde{K}^* \Bigg)^{\Gamma} & = \frac{1}{1 - q} \sum_{\ell = 0}^{n(k + 1) - 1} \sum_{a, b = 1}^{n(k + 1)} K_a \widetilde{K}^*_b (\tilde{q}^\ell R_a \otimes R_b^*)^{\mathbb{Z}_{n(k + 1)}} \\
& = \frac{1}{1 - q} \sum_{b \geq a} \tilde{q}^{b - a} \widetilde{K}_a \widetilde{K}_b^* + \frac{1}{1 - q} \sum_{b < a} \tilde{q}^{b - a + n(k + 1)} \widetilde{K}_a \widetilde{K}_b^* \\
& = \frac{1}{1 - q}\sum_{c_K(j) \geq c_K(i)} \frac{x_i}{x_j} + \frac{q}{1 - q} \sum_{c_K(j) < c_K(i)} \frac{x_i}{x_j}. 
\end{split}
\end{equation}
Taking $S^\bullet$ of (minus) this leads to the contribution to the integrand  
\begin{equation}
\prod_{i \geq j} \varphi_q(x_i/x_j) \prod_{i < j}\varphi_q(qx_i/x_j) = \prod_{i < j}(1 - x_i/x_j)^{-1} \prod_{i \neq j} \varphi_q(x_i/x_j). 
\end{equation}
The last factor is the usual contribution of the vector multiplet, and what remains is the contribution of the vector multiplet to the defect operator, due to the singular boundary condition on the gauge fields. 

For the contribution of the adjoint chiral, the result is:
\begin{equation}
\begin{split}
\Bigg(\frac{\hbar^{-1} \tilde{q}}{1 - \tilde{q}}\widetilde{K} \widetilde{K}^* \Bigg)^{\Gamma} & = \frac{\hbar^{-1}}{1 - q} \sum_{\ell = 0}^{n(k + 1) - 1} \sum_{a, b = 1}^{n(k + 1)} K_a \widetilde{K}^*_b (\tilde{q}^{\ell + 1} R_a \otimes R_b^*)^{\mathbb{Z}_{n(k + 1)}} \\
& = \frac{\hbar^{-1}}{1 - q} \sum_{b > a} \tilde{q}^{b - a} \widetilde{K}_a \widetilde{K}_b^* + \frac{\hbar^{-1}}{1 - q} \sum_{b \leq a} \tilde{q}^{b - a + n(k + 1)} \widetilde{K}_a \widetilde{K}_b^* \\
& = \frac{\hbar^{-1}}{1 - q}\sum_{c_K(j) > c_K(i)} \frac{x_i}{x_j} + \frac{\hbar^{-1}q}{1 - q} \sum_{c_K(j) \leq c_K(i)} \frac{x_i}{x_j}. 
\end{split}
\end{equation}
Taking $S^\bullet$ of this leads to the contribution 
\begin{equation}
\prod_{i > j} \varphi_q(\hbar^{-1} x_i/x_j)^{-1} \prod_{i \leq j}\varphi_q(tx_i/x_j)^{-1} = \prod_{i > j}(1 - \hbar^{-1}x_i/x_j)^{-1} \prod_{i \neq j} \varphi_q(tx_i/x_j)^{-1}.
\end{equation}

The fundamental and antifundamental chiral multiplets contribute
\begin{equation}
\prod_{c_K(i) > c_+(\alpha)}(1 - x_i/m_\alpha^+) \prod_{c_K(i) < c_-(\alpha)} (1 - x_i/m_\alpha^-) \, \prod_{i = 1}^k \prod_{\alpha = 1}^n \frac{\varphi_q(qx_i/m_\alpha^-)}{\varphi_q(x_i/m_\alpha^+)}. 
\end{equation}
From the definitions of the coloring functions, one sees that the condition $c_K(i) > c_+(\alpha)$ is equivalent to $\lambda_i > \alpha$, and likewise $c_K(i) < c_-(\alpha)$ is equivalent to $\lambda_i < \alpha$. 

\subsubsection{}

It follows the partition function of the theory in the presence of the defect labeled by the fixed point $\lambda$ (viewed as a partition) is given by the integral 
\begin{equation} \label{stabint}
\psi_\lambda(q, \hbar, z, m_\alpha^\pm) = \int [dx] \prod_{i = 1}^k x_i^{-\frac{\log(z)}{\log (q)}} \prod_{ 1 \leq i \neq j \leq k} \frac{\varphi_q(x_i/x_j)}{\varphi_q(tx_i/x_j)} \prod_{\alpha = 1}^n \prod_{i = 1}^k \frac{\varphi_q(qx_i/m_\alpha^-)}{\varphi_q(x_i/m_\alpha^+)} \frac{\text{Stab}^K_\lambda(x)}{\Delta_\hbar(x)},
\end{equation}
where 
\begin{equation}\label{sl}
\text{Stab}^K_\lambda(x) = \text{Sym}\prod_{i < j} \frac{1 - \hbar^{-1}x_i/x_j}{1 - x_i/x_j} \prod_{i = 1}^k p_{\lambda_i}(x_i), 
\end{equation}
and  
\begin{equation}\label{psl}
p_{\lambda_i}(x_i) = \prod_{\alpha < \lambda_i} \Big(1 - \frac{x_i}{m_\alpha^+} \Big) \prod_{\alpha > \lambda_i} \Big(1 - \frac{x_i}{m_\alpha^-} \Big). 
\end{equation}
We have introduced the universal denominator 
\begin{equation}
\Delta_\hbar(x) = \prod_{i \neq j} (1 - \hbar^{-1}x_i/x_j) .
\end{equation}
We are free to symmetrize, since we integrate against a measure symmetric in $x$ variables.
\subsubsection{}
The classes $\text{Stab}^K_\lambda \in K_T(X)$ indexed by fixed points $\lambda$ and defined in eqns. \eqref{sl} and \eqref{psl} are the (K-theoretic) stable envelopes for the variety $X$. They share many features with the stable basis known to exist for Nakajima varieties. K-theoretic stable envelopes are defined in \cite{okounkov15} as the unique (after making certain choices, see below) classes satisfying certain support, degree, and normalization conditions and it may be verified that the classes written above satisfy the relevant conditions for $X$. We present some of the details of this later on, in the elliptic context (from which $K$-theoretic statements follow by taking limits). 

Write
\begin{equation}
\begin{split}
m_\alpha^+ & = a_\alpha \\
m_\alpha^- & = a_\alpha \hbar^{\mu_\alpha},
\end{split}
\end{equation}
where $a_\alpha$ is the position on the cylinder of the vertex operator, as before. Comparing to \cite{afo17}, when $\mu_\alpha = 1$ for each $\alpha$, these formulas coincide with the known results for the fundamental vertex operators. This is one way to anticipate that we should identify $\mu_\alpha$ with the highest weight of the $\mathscr{U}_\hbar(\widehat{\mathfrak{sl}_2})$ Verma module inserted at $a_{\alpha}$, where $\mu_\alpha$ is taken to be a generic complex number. 
\subsection{qKZ Equation}
We now present a direct proof 
that \eqref{stabint} produces an integral solution to the qKZ equation for $q$-conformal blocks of $\mathscr{U}_\hbar(\widehat{\mathfrak{sl}_2})$ with $n$ Verma module operator insertions on the cylinder $\mathbb{C}^\times$. This is the analog of the result of \cite{ao17}. While we used integral formulas rather than geometric language in the previous section, it is easy to verify that the integral \eqref{stabint} represents a quasimap vertex function with (localized) descendent insertion given by the stable envelope, which is the setup considered in \cite{ao17}.

We begin by recalling some background on qKZ, useful facts about stable envelopes, and setting up our notations.

\subsubsection{}
The qKZ equations are intimately connected to integrable lattice models and spin chains. The connection of gauge theories to integrable spin systems was first observed in \cite{ns09} on the level of Bethe equations, which we recall here. 

The starting point is the stationary phase asymptotics of \eqref{partitionfn} as $q \to 1$ with all other parameters fixed, corresponding to removing the $\Omega$-background and studying the theory on flat space. The integrand behaves asymptotically as $e^{\widetilde{W}_{\text{eff}}/\log q}$ where $\widetilde{W}_{\text{eff}}(x_i)$ is the effective twisted superpotential of the gauge theory, expressed as a function of scalars $x_i$ parameterizing the Coulomb branch. The partition function is then dominated by critical points of $\widetilde{W}_{\text{eff}}$, corresponding to supersymmetric vacua of the theory. At the critical points, $x_i$ solve the following Bethe equations:
\begin{equation}
\prod_{\alpha = 1}^n \frac{1 - x_i/m_\alpha^+}{1 - x_i/m_\alpha^-} = \hbar^{k - 1} z \prod_{j (\neq i)}\frac{1 - \hbar^{-1}x_j/x_i}{1 - \hbar x_j/x_i}. 
\end{equation}
There is one equation for each $i = 1, \dots, k$. The Bethe equations are those of the XXZ spin chain, with  complex spins, studied in \cite{Chen_2011}. 

The space of supersymmetric vacua is the $T$-equivariant quantum K-theory of $X$. The latter is defined as the ring of symmetric Laurent polynomials of $x_i$ variables, the Chern roots of the tautological bundle $V$, modulo the relations defined by the Bethe equations. The fact that these arise from the stationary phase asymptotics of \eqref{partitionfn}, which is itself a certain quasimap count, makes manifest that the product in equivariant quantum K-theory takes into account the contributions of rational curves in $X$. 

As pointed out in \cite{ns09}, the presence of Bethe equations suggests the presence of an underlying quantum symmetry algebra, in this case the quantum affine algebra $\mathscr{U}_\hbar(\widehat{\mathfrak{sl}_2})$---the fact that the spins in the Bethe equations are generic and complex is one way to understand why one expects the theory under consideration to realize Verma modules. 

A powerful way of constructing actions of such algebras starts from the geometric construction of an $R$-matrix, a point of view developed extensively starting from \cite{mo12}. In \cite{ao17} it was explained how gauge theory and geometry may be used to construct explicit integral solutions of the qKZ equation associated to such $R$-matrices; it is this construction that we extend to our current example in what follows. 

The key ingredient in the geometric construction of an $R$-matrix are the stable envelopes.

\subsubsection{}
Stable envelopes depend on a choice of a chamber of real mass parameters for the torus $A$ (as well as a polarization \cite{okounkov15, okounkov20}---our polarization is fixed to \eqref{Tsqrt} throughout). The chamber giving rise to the explicit formula written above is $|a_1| < |a_2| <  \dots < |a_n|.$ Stable envelopes in each chamber $\mathfrak{C}$ provide a natural basis in $K_T(X)$ for realizing the geometric action of quantum groups; in particular, the transition matrix between stable envelopes in two different chambers is by definition the geometric $R$-matrix: 
\begin{equation}
\text{Stab}_{\mathfrak{C}'} = R_{\mathfrak{C} \to \mathfrak{C}'} \text{Stab}_{\mathfrak{C}}. 
\end{equation}
It follows quite generally that such an $R$-matrix automatically solves the quantum Yang-Baxter equation, see section 9 in \cite{okounkov15}. For our stable envelopes, the geometric $R$-matrix can be identified with the $\mathscr{U}_\hbar(\widehat{\mathfrak{sl}_2})$  $R$-matrix appearing in textbooks \cite{kzbook} acting in a tensor product of evaluation Verma modules (some explicit formulas to this effect will be presented in later sections). 
\subsubsection{}
To keep the notation light we restrict to the case of the qKZ equation associated to $n = 2$ insertions on the cylinder, however the argument may be extended to general $n$ in a straightforward fashion. Then, on the geometric side, there are only two possible chambers, namely $|a_1| < |a_2|$ and $|a_2| < |a_1|$. 
 
There are $k + 1$ $T$-fixed points on $X$ indexed by an integer $0 \leq \ell \leq k$. Denote the stable envelopes in each chamber by $\text{Stab}^{12}_\ell$ and $\text{Stab}^{21}_\ell$. The transition matrix between them is by definition the geometric $R$-matrix: 
\begin{equation} \label{rmatrixdef}
\text{Stab}^{21}_\ell = \sum_{\ell'}R_{12}(a_1/a_2)_{\ell \ell'} \text{Stab}^{12}_{\ell'}. 
\end{equation}
It is a basic geometric fact, following readily from definitions, that $R_{12}(a_1/a_2)^{-1} = R_{21}(a_2/a_1)$. \eqref{rmatrixdef} is understood in $K_T(X)$, however the following stronger statement is true.  The stable envelopes in each chamber are given by the explicit formulas 
\begin{align}
\text{Stab}^{12}_\ell(x, a) & = \text{Sym} \prod_{i < j} \frac{1 - \hbar^{-1}x_i/x_j}{1 - x_i/x_j} \prod_{i \leq \ell} \Big( 1 - \frac{x_i}{a_1} \Big) \prod_{i > \ell} \Big( 1 - \frac{x_i}{\hbar^{\mu_2} a_2} \Big) \\
\text{Stab}^{21}_\ell(x, a) & = \text{Sym} \prod_{i < j} \frac{1 - \hbar^{-1}x_i/x_j}{1 - x_i/x_j} \prod_{i \leq k - \ell} \Big( 1 - \frac{x_i}{a_2} \Big) \prod_{i > k - \ell} \Big(1 - \frac{x_i}{\hbar^{\mu_1}a_1} \Big). 
\end{align}
As classes in $K_T(X)$, these are regarded as polynomials in the variables $x_i$ modulo a certain ideal. Regard them instead as honest polynomials in $x_i$, corresponding geometrically to a pushforward from $X$ to the ambient quotient stack. It is clear from the above formulas that each stable envelope is a symmetric polynomial in $x_1, \dots, x_k$ of degree at most one in each variable. The vector space of such polynomials is of dimension $k + 1$, which is the same as the number $T$-fixed points on $X$ and the number of stable envelopes in each chamber. Moreover, stable envelopes are linearly independent as polynomials (this follows immediately from the support condition), so \eqref{rmatrixdef} holds as an identity among polynomials in $x$-variables. This is a useful simplifying feature in the following arguments. 
\subsubsection{}
The qKZ equations for a function $\psi(a_1, a_2)$ valued in 
\begin{equation}
\Big( \mathscr{V}_{\mu_1}(a_1) \otimes \mathscr{V}_{\mu_2}(a_2) \Big)_{\text{weight} = \frac{\mu_1 + \mu_2}{2} - k}
\end{equation}
read 
\begin{align}
\psi(qa_1, a_2) & = (z^h \otimes 1) R_{12}(a_1/a_2) \psi(a_1, a_2) \\
\psi(a_1, qa_2) & = R_{21}(qa_2/a_1) (1 \otimes z^h) \psi(a_1, a_2).
\end{align}
The operator $z^h$ acts on a vector $v_w$ of weight $w$  by recording the weight:
\begin{equation}
z^h(v_w) = z^w v_w 
\end{equation}
Since $\mathfrak{sl}_2$ has rank 1, we need only one $z$-variable. In general, there are $\text{rk}( \mathfrak{g})$-many $z$-variables and they correspond to the action of the Cartan $\mathfrak{h} \subset \mathfrak{g}$. 

Let $f$ be the lowering operator of $\mathfrak{sl}_2$, and $v_1$, $v_2$ be the highest weight vectors of Verma modules at $a_1$ and $a_2$. Expand 
\begin{equation}
\psi(a_1, a_2) = \sum_{\ell = 0}^k \psi_\ell (a_1, a_2) f^{k - \ell} v_1 \otimes f^\ell v_2.
\end{equation}
In terms of these components, the qKZ equations are 
\begin{align}
\psi_\ell(qa_1, a_2) & = \sum_{\ell'} z^{\frac{\mu_1}{2} - (k - \ell)} R_{12}(a_1/a_2)_{\ell \ell'} \psi_{\ell'}(a_1, a_2) \\
\psi_{\ell}(a_1, qa_2) & = \sum_{\ell'} R_{21}(qa_2/a_1)_{\ell \ell'} z^{\frac{\mu_2}{2} - \ell'} \psi_{\ell'}(a_1, a_2).
\end{align}
\subsubsection{}
Next, write 
\begin{equation}
\psi_\ell(a_1, a_2) = \exp{\frac{\log(z)(\mu_1 \log(a_1) + \mu_2 \log(a_2))}{2\log(q)}} \times I_\ell(a_1, a_2). 
\end{equation}
Then the claim is that if $I_\ell(a_1, a_2)$ is given by the integral
\begin{equation}
 \int_\gamma \prod_{j = 1}^k \frac{dx_j}{2\pi i x_j} \prod_i x_i^{-\frac{\log(z)}{\log(q)}} \prod_{i \neq j } \frac{\varphi_q(x_i/x_j)}{\varphi_q(\hbar^{-1} x_i/x_j)} \prod_{i = 1}^k \frac{\varphi_q(qx_i/\hbar^{\mu_1} a_1) \varphi_q(qx_i/\hbar^{\mu_2} a_2)}{\varphi_q(x_i/a_1) \varphi_q(x_i/a_2)} \text{Stab}^{12}_\ell(x, a) 
\end{equation}
then $\psi(a_1, a_2)$ solves the above qKZ equations. The only property we require of the contour $\gamma$ is that it is invariant under $x_i \to qx_i$ for each $i$. 

This is proven as follows. Because $\text{Stab}^{12}_\ell$ is integrated against a measure symmetric in $x_i$ variables, one can drop the symmetrization and the integrand, neglecting the monomial piece, becomes the product of the factors 
\begin{equation} \label{gaugeqkz}
\prod_{i < j} \frac{\varphi_q(q x_i/x_j) \varphi_q(x_j/x_i)}{\varphi_q(q\hbar^{-1} x_i/x_j) \varphi_q(\hbar^{-1} x_j/x_i)} 
\end{equation}
and 
\begin{equation} \label{matterqkz}
\prod_{i \leq \ell} \frac{\varphi_q(q x_i/\hbar^{\mu_2} a_2)}{\varphi_q(qx_i/a_1)} \prod_{i > \ell} \frac{\varphi_q(x_i/\hbar^{\mu_2} a_2)}{\varphi_q(x_i/a_1)} \prod_i \frac{\varphi_q(qx_i/\hbar^{\mu_1}a_1)}{\varphi_q(x_i/a_2)}. 
\end{equation}
Now consider the action of the shift $a_1 \to q a_1$. Using the invariance of the integration contour, shift also $x_i \to q x_i $ for $i > \ell$. This gives rise to an overall factor of $z^{-(k - \ell)}$ from the monomial term in the integrand. \eqref{matterqkz} becomes 
\begin{equation}
\prod_{i = 1}^k \frac{\varphi_q(qx_i/\hbar^{\mu_1} a_1) \varphi_q(qx_i/\hbar^{\mu_2} a_2)}{\varphi_q(x_i/a_1) \varphi_q(x_i/a_2)} \prod_{i \leq \ell} (1 - x_i/\hbar^{\mu_1} a_1) \prod_{i > \ell} (1 - x_i/a_2). 
\end{equation}
Likewise \eqref{gaugeqkz} becomes 
\begin{equation}
\prod_{i \neq j} \frac{\varphi_q(x_i/x_j)}{\varphi_q(\hbar^{-1} x_i/x_j)} \prod_{i < j \leq \ell} \frac{1 - \hbar^{-1} x_i/x_j}{ 1 - x_i/x_j} \prod_{i \leq \ell, j > \ell} \frac{1 - \hbar^{-1}x_j/x_i}{1 - x_j/x_i} \prod_{\ell < i < j } \frac{1 - \hbar^{-1} x_i/x_j}{1 - x_i/x_j}.  
\end{equation}
Gathering terms and once again using the freedom to permute the $x_i$, we  recognize the new insertion as 
\begin{equation}
\prod_{i < j} \frac{1 - \hbar^{-1}x_i/x_j}{1 - x_i/x_j} \prod_{i \leq k - \ell} \Big(1 - \frac{x_i}{a_2} \Big) \prod_{ i > k - \ell} \Big(1 - \frac{x_i}{\hbar^{\mu_1} a_1} \Big)
\end{equation}
which becomes $\text{Stab}^{21}_\ell$ upon symmetrization. So far we have established that $I_\ell(qa_1, a_2)$ is equal to
\begin{equation}
z^{- (k - \ell)} \int_\gamma \prod_{j = 1}^k \frac{dx_j}{2\pi i x_j} \prod_i x_i^{-\frac{\log(z)}{\log(q)}} \prod_{i \neq j} \frac{\varphi_q(x_i/x_j)}{\varphi_q(\hbar^{-1}x_i/x_j)} \prod_{i = 1}^k \frac{\varphi_q(qx_i/\hbar^{\mu_1} a_1) \varphi_q(qx_i/\hbar^{\mu_2} a_2)}{\varphi_q(x_i/a_1) \varphi_q(x_i/a_2)} \text{Stab}^{21}_\ell(x, a). 
\end{equation}
Now using \eqref{rmatrixdef} under the integral and passing from $I_\ell$ to $\psi_\ell$, it follows that 
\begin{equation}
\psi_\ell(qa_1, a_2) = \sum_{\ell'} z^{\frac{\mu_1}{2} - (k - \ell)} R_{12}(a_1/a_2)_{\ell \ell'} \psi_{\ell'}(a_1, a_2)
\end{equation}
which is nothing but the qKZ equation. 
\subsubsection{}
The argument for $a_2 \to qa_2$ is similar. First write 
\begin{equation}
\text{Stab}^{12}_\ell(x, a_1, qa_2) = \sum_{\ell'} R_{21}(qa_2/a_1)_{\ell \ell'} \text{Stab}^{21}_{\ell'}(x, a_1, qa_2)
\end{equation}
under the integral, and focus on each term in the sum. Proceeding as before, the factor \eqref{gaugeqkz} in the integrand is the same, and now is multiplied with 
\begin{equation}
\prod_{i \leq k - \ell'} \frac{\varphi_q(qx_i/\hbar^{\mu_1} a_1) \varphi_q(x_i/\hbar^{\mu_2} a_2)}{\varphi_q(x_i/a_1) \varphi_q(x_i/a_2)} \prod_{i > k - \ell'} \frac{\varphi_q(x_i/\hbar^{\mu_1}a_1)}{\varphi_q(x_i/a_1)}. 
\end{equation}
Now shifting $x_i \to qx_i$ for $i > k - \ell'$, \eqref{gaugeqkz} gives rise to a similar piece as before, while the above gives rise to 
\begin{equation}
\prod_{i = 1}^k \frac{\varphi_q(qx_i/\hbar^{\mu_1} a_1) \varphi_q(qx_i/\hbar^{\mu_2} a_2)}{\varphi_q(x_i/a_1) \varphi_q(x_i/a_2)} \prod_{i > k - \ell'} (1 - x_i/a_1) \prod_{i \leq k - \ell'} (1 - x_i/\hbar^{\mu_2} a_2).
\end{equation}
Again, relabeling the $x_i$ variables the insertion is recognized as 
\begin{equation}
\prod_{i < j} \frac{1 - \hbar^{-1} x_i/x_j}{1 - x_i/x_j} \prod_{i \leq \ell'} \Big(1 - \frac{x_i}{a_1} \Big) \prod_{i > \ell'} \Big(1 - \frac{x_i}{\hbar^{\mu_2} a_2} \Big)
\end{equation}
which becomes $\text{Stab}^{12}_{\ell'}$ upon symmetrization. Because we shifted $\ell'$ $x_i$ variables by $q$, we obtain a factor $z^{-\ell'}$ from the monomial piece of the integrand, and combining all factors we obtain 
\begin{equation}
\psi_\ell(a_1, qa_2) = \sum_{\ell'} R_{21}(qa_2/a_1)_{\ell \ell'} z^{\frac{\mu_2}{2} - \ell'} \psi_{\ell'}(a_1, a_2).
\end{equation}
This is what we wanted to show. It is not difficult to extend this reasoning to $n > 2$, and we leave this to the interested reader. 

\subsection{Elliptic Stable Envelopes, Support, Poles and Monodromy} \label{ellipticstabandpoles}
The stable envelopes admit an uplift to equivariant elliptic cohomology $\text{Ell}_T(X)$ \cite{ao16}. It has been explained above that the stable envelopes in K-theory are interpreted in gauge theory terms as a distinguished class of line defects. The elliptic envelopes arise as boundary conditions in the three dimensional gauge theories, described by coupling the three dimensional theory in the $\mathbb{C} \times S^1$ bulk to a two-dimensional theory on the $T^2$ boundary at infinity. At the level of integral formulas for partition functions, the boundary theory contributes via its elliptic genus, leading to additional insertions of theta functions into the integral. 

The elliptic envelopes realize geometrically the ``pole subtraction matrices'' converting the solutions of qKZ analytic in $z = e^{-\zeta}$ (the K\"{a}hler parameter) to those analytic in the $a_\alpha$ variables. The latter are the honest $q$-conformal blocks (for more background on this see \cite{afo17}). In the present section, we present the formulas for elliptic stable envelopes, demonstrate that they satisfy the support condition, and explain the pole subtraction mechanism concretely in simple examples. Understanding pole subtraction immediately allows one to deduce that the monodromy of the qKZ equation is controlled by elliptic R-matrices. 
\subsubsection{}
Define the theta function 
\begin{equation}
\vartheta(x) = (x^{1/2} - x^{-1/2})\varphi_q(qx)\varphi_q(q/x).
\end{equation}
For our variety $X$, the elliptic stable envelopes are given explicitly by the formula 
\begin{equation} \label{ellipticstabdef}
\text{Stab}^{Ell}_\lambda(x) = \frac{1}{\Theta(x)} \text{Sym} \prod_{i < j} \frac{\vartheta(\hbar^{-1}x_i/x_j)}{\vartheta(x_i/x_j)} \prod_{i = 1}^k P_{\lambda_i}(x_i) 
\end{equation}
where 
\begin{equation}
\Theta(x)  = \prod_{i = 1}^k \prod_{\alpha = 1}^n \vartheta(x_i/m_\alpha^-) 
\end{equation}
and
\begin{equation} \label{plambda}
P_{\lambda_i}(x_i)  = \vartheta\Big( \frac{z x_i}{m_{\lambda_i}^-} \hbar^{k + 1 - 2i - \sum_{\alpha < \lambda_i} \mu_\alpha} \Big) \prod_{\alpha < \lambda_i} \vartheta(x_i/m_\alpha^+)\prod_{\alpha > \lambda_i}\vartheta(x_i/m_\alpha^-). 
\end{equation}
Elliptic stable envelopes (associated to a given chamber of $a$-variables) are defined in \cite{ao16} as the unique elliptic cohomology classes satisfying analogs of the degree, support, and normalization conditions appropriate in the elliptic cohomology setting. Equivariant elliptic cohomology classes are by definition sections of appropriate line bundles on the scheme $\text{Ell}_T(X)$. We will not attempt to review basics on elliptic cohomology here, and refer to section 2 of \cite{ao16} for that. What is important in practice is that the classes relevant for us can just be viewed concretely as elliptic functions of the $K$-theoretic Chern roots of tautological bundles, and this is how the formulas above are to be understood.

That the appropriate normalization condition is satisfied is a trivial consequence of the explicit formulas above. Likewise, the analog of the degree condition in elliptic cohomology is a specification of the line bundle of which the stable envelopes are a section; in practice this just amounts to specifying factors of automorphy under $q$-shifts of $x_i$ and $a_\alpha$ variables. It is again straightforward to verify the analog of the degree condition of \cite{ao16} in the present setting from the explicit formulas, using $\vartheta(qx) = (-\sqrt{q} x)^{-1} \vartheta(x)$. 

While it is a minor point, we should point out that to literally match the normalization and degree conditions of \cite{ao16} (modulo the small adjustments necessary since $X$ is not a Nakajima variety that also appeared in section \ref{quivergauge}), the $\Theta(x)$ in the denominator of \eqref{ellipticstabdef} should be removed, and \eqref{plambda} should be divided by $\vartheta(z\hbar^{k + 1 - 2i -\sum_{\alpha \leq \lambda_i} \mu_\alpha})$\footnote{Also, a few other factors should be inserted which depend only on $\hbar, \mu_\alpha$, whose role is to cancel parts of the normal bundle to each fixed point which have trivial $A$-weights. We are grateful to the authors of \cite{zhouetal2024} for pointing this out to us.} The reason for the slightly different choices here is that we will be most interested in applications to integral formulas, compare also to the discussion around eq. 3.31 of \cite{afo17}. 

The support condition asserts that, for each connected component of $X^A$, the corresponding elliptic stable envelope is supported on its full attracting locus for the given chamber $\mathfrak{C}$ of $a$-variables. This condition is the least immediate and requires verifying a certain vanishing property of the elliptic stable envelopes that we spell out presently; again we focus on the case of $n = 2$ $a$-variables to keep notations simple, but there is no essential difficulty extending the argument presented to $n > 2$. 
\subsubsection{}
Before discussing the calculation, we remind the reader of a few simple facts. The explicit formula for elliptic stable envelopes becomes, in the $n = 2$ case, 
\begin{equation}
\text{Stab}^{Ell}_\ell(x) = \frac{1}{\Theta(x)} \text{Sym} \prod_{i < j} \frac{\vartheta(\hbar^{-1} x_i/x_j)}{\vartheta(x_i/x_j)}  P_\ell(x)
\end{equation}
where
\begin{equation}
P_\ell(x)=
\prod_{i \leq \ell } \vartheta\Big( \frac{x_iz\hbar^{k + 1 - 2i - \mu_1 - \mu_2}}{a_2} \Big) \vartheta(x_i/a_1) \prod_{ i > \ell} \vartheta\Big( \frac{x_i z \hbar^{k + 1 - 2i - \mu_1}}{a_1} \Big) \vartheta(x_i/\hbar^{\mu_2} a_2)
\end{equation}
where $\Theta(x)$ is as above with $n = 2$. Connected components of the $A$-fixed locus are labeled by the integer $0 \leq \ell \leq k$. The connected component $F_j$ of the fixed locus is in the full attracting set $\text{Attr}(F_\ell)$ if and only if $j \geq \ell$. It suffices to verify that the restriction of $\text{Stab}^{Ell}_\ell$ to the fixed point $j$ vanishes unless $j \geq \ell$. 
\subsubsection{}
The strategy is to use the contour integrals to facilitate the computation. Due to the factor of automorphy under $x_i \to q x_i$ of elliptic stable envelopes, integrals of the form 
\begin{equation}
I_\ell = \int_\gamma \prod_{j = 1}^k \frac{dx_j}{2\pi i x_j} \text{Stab}^{Ell}_\ell(x, a) \prod_{i \neq j} \frac{\varphi_q(x_i/x_j)}{\varphi_q(\hbar^{-1} x_i/x_j)} \prod_{i, \alpha} \frac{\varphi_q(qx_i/\hbar^{\mu_\alpha} a_\alpha)}{\varphi_q(x_i/a_\alpha)}
\end{equation}
are linear combinations of vertex functions, with coefficients given by the restrictions of $\text{Stab}^{Ell}_\ell$ to $T$-fixed points in $X$. In the notation above, 
\begin{equation} \label{ellstabint}
I_\ell = \sum_{p \in X^T} \text{Stab}^{Ell}_\ell \eval_p Z_p. 
\end{equation}
It is clear from the above formula that the support of the elliptic stable envelope is determined by which fixed points $p$ may have nonzero contributions, since the $Z_p$ are linearly independent. As discussed at length above, concretely this may be understood by analyzing the poles of the integrand. In the chamber $|z| < 1$, it suffices to look at the sequences of poles accumulating at $|x| \to \infty$. 

Before proceeding to the details, we remark that any superficial pole with $x_i/x_j \in q^{\mathbb{Z}}$ for $i \neq j$ turns out to have vanishing residue. This is most clear in the original expression \eqref{ellstabint}: elliptic stable envelopes introduce no new poles for $|x| \to \infty$ (those associated to $\Theta(x)$ in the denominator only appear for $|x| \to 0$), and the factor $\varphi_q(x_i/x_j)$ in the numerator vanishes on any such point, ensuring that no such points contribute to the sum over residues. 
\subsubsection{}
The strategy for analyzing the poles is as follows. As usual, we drop the symmetrization in $\text{Stab}^{Ell}_\ell$ since the integration measure is symmetric in $x_i$. We have that
\begin{equation} \label{ellstabgauge}
\prod_{i < j} \frac{\vartheta(\hbar^{-1} x_i/x_j)}{\vartheta(x_i/x_j)} \frac{\varphi_q(x_i/x_j)\varphi_q(x_j/x_i)}{\varphi_q(\hbar^{-1}x_i/x_j) \varphi_q(\hbar^{-1}x_j/x_i)}=
(\text{prefactor}) \times \prod_{i < j} (1 - x_j/x_i) \frac{\varphi_q(q\hbar x_j/x_i)}{\varphi_q(\hbar^{-1}x_j/x_i)}
\end{equation}
and that
\begin{equation} \label{ellstabmatter}
\frac{1}{\Theta} \prod_i \frac{\varphi_q(qx_i/\hbar^{\mu_1}a_1) \varphi_q(qx_i/\hbar^{\mu_2}a_2)}{\varphi_q(x_i/a_1)\varphi_q(x_i/a_2)} \prod_{i \leq \ell }\vartheta(x_i/a_1) \prod_{i > \ell} \vartheta(x_i/\hbar^{\mu_2}a_2)
\end{equation}
equals a prefactor times 
\begin{equation}
\prod_{i \leq \ell} \frac{\varphi_q(qa_1/x_i)}{\varphi_q(x_i/a_2) \varphi_q(\hbar^{\mu_1} a_1/x_i) \varphi_q(\hbar^{\mu_2} a_2/x_i)} \prod_{i > \ell} \frac{\varphi_q(qx_i/\hbar^{\mu_2} a_2)}{\varphi_q(x_i/a_1) \varphi_q(x_i/a_2) \varphi_q(\hbar^{\mu_1} a_1/x_i)}. 
\end{equation}
In the prefactors, we have hidden various inessential terms such as constants, signs, and powers of $x$ which do not affect the singularity structure. The poles of the integrand are determined by the poles in the expressions \eqref{ellstabgauge} and \eqref{ellstabmatter}.

The only sequence of poles in \eqref{ellstabgauge} and \eqref{ellstabmatter} in the variable $x_1$, which accumulates at $|x| \to \infty$, is 
\begin{equation}
x_1 = a_2 q^{-n_1}
\end{equation}
for integers $n_1 \geq 0$, coming from the denominator in \eqref{ellstabmatter}. Bearing in mind that $x_2/x_1 \notin q^{\mathbb{Z}}$ to have nonvanishing residue, the only sequence of poles in $x_2$ accumulating at $|x| \to \infty$ is 
\begin{equation}
x_2 = \hbar x_1 q^{-n_2} = \hbar a_2 q^{-n_1 - n_2}
\end{equation}
coming from the denominator of \eqref{ellstabgauge}. Iterating, the only sequences of poles accumulating at $|x| \to \infty$ in the variables $x_i$, for $i \leq \ell$, are 
\begin{equation}
x_i = \hbar^{i - 1}a_2 q^{- (n_1 + \dots + n_i)}.
\end{equation}
for integers $n_1, \dots, n_i \geq 0$. So far, we have made no choices, as these sequences of poles are required simply by asking that the integral is nonzero. For the variables $x_i$ with $i > \ell$, there is some choice but it is quite restrictive. Inspecting the denominators in \eqref{ellstabmatter} and \eqref{ellstabgauge}, the only possibility is to continue growing the ``tail'' attached to $a_2$ by picking poles from \eqref{ellstabgauge}, or to start growing a new ``tail'' at $a_1$ by choosing one of the variables to be at a zero of $\varphi_q(x_i/a_1)$ in \eqref{ellstabmatter}. 

Up to relabeling of the $x_i$ variables, the most general sequences of poles which contribute are of the form 
\begin{align}
x_i & = \hbar^{i - 1} a_2 q^{-\lambda_{j - i + 1}}, & i \leq j, \\
x_i & = \hbar^{i - j - 1 }a_1 q^{-\mu_{k - i + 1}},  & i > j. 
\end{align}
In the above, $\lambda_i$ and $\mu_i$ are two partitions, the first of length at most $j$ and the second of length at most $k - j$, and $j$ is an integer such that $\ell \leq j \leq k$. It is clear that these sequences correspond precisely to the fixed points $j$ with $j \geq \ell$, which are those corresponding to the components $F_j$ in $\text{Attr}(F_\ell)$. This establishes the support condition for $\text{Stab}^{Ell}$. 
\subsubsection{}
In the simplest example, namely the $k=1$ case, it is simple to see directly how the elliptic stable envelopes eliminate poles of the partition function in the $a$-variables. In this case, fixed points are labeled by an integer $i$ between $1$ and $n$. Consider, for a pair of fixed points $(i, j)$, the integral  
\begin{equation}
 (\mathscr{F})\indices{^i_j}(a_\alpha; z) = \int_\gamma \frac{dx}{2\pi i x} x^{-\frac{\log z}{\log q}} \mathscr{P}^i(x; a, z) \prod_{\alpha = 1}^n \frac{\varphi(qx/\hbar^{\mu_\alpha} a_\alpha)}{\varphi(x/a_\alpha)} \text{Stab}^K_j(x).    
\end{equation}
The insertion $\mathscr{P}^i(x)$ is given by 
\begin{multline}
 \mathscr{P}^i(x) = \exp{\frac{ \log(a_i) \log(z^{-1} \hbar^{\sum_{\alpha < i} \mu_\alpha})}{\log q} - \sum_{\alpha < i} \frac{\log(a_\alpha) \log(\hbar^{\mu_\alpha})}{\log q} } \times \\ \frac{\vartheta(xz\hbar^{-\sum_{\alpha \leq i} \mu_\alpha}/a_i)}{\vartheta(z \hbar^{-\sum_{\alpha \leq i} \mu_\alpha})} \frac{\prod_{\alpha < i} \vartheta(x/a_\alpha) \prod_{\alpha > i} \vartheta(x/\hbar^{\mu_\alpha} a_\alpha)}{\prod_{\alpha = 1}^n \vartheta(x/\hbar^{\mu_\alpha}a_\alpha)} \exp{\frac{\log(x)\log(z)}{\log q}}. 
\end{multline}
Up to normalizations, this is the elliptic stable envelope associated to the fixed point $i$. The extra factors are present to ensure the insertion is invariant with respect to $q$-shifts of all variables: 
\begin{equation}
\mathscr{P}^i(x; a_1, \dots, a_n; z) = \mathscr{P}^i(qx, a_1, \dots, a_n; z) = \mathscr{P}^i(x; a_1, \dots, qa_\alpha, \dots, a_n; z) = \mathscr{P}^i(x; a_1, \dots, a_n; qz). 
\end{equation}
In particular, this ensures that the integral with the insertion of $\mathscr{P}^i(x)$ for any $i$ solves the same $q$-difference equation in the $a$-variables as the integral without the insertion, namely, it solves the qKZ equation. 

Contour integrals of the type above develop singularities when a pair of poles on either side of the contour $\gamma$ coalesce as parameters are varied. We will show that the integral with the insertion of $\mathscr{P}^i$ is analytic in $a$-variables in the chamber $|a_1| < \dots < |a_n|$, corresponding to the choice of chamber for the elliptic stable envelopes. The contour $\gamma$ is chosen to separate the sequences of poles of the integrand, located at:  
\begin{align*}
x & = a_\alpha q^{-n}, & \alpha \geq i, \; n \geq 0 \\
x & = \hbar^{\mu_\alpha} a_\alpha q^n,  & \alpha \leq i, \; n \geq 0. 
\end{align*}
The first set of sequences accumulate to $|x| \to \infty$, while the second accumulate to $|x| \to 0$. 

Two poles from opposite sides of the integration contour may collide when the equation 
\begin{equation}
\frac{a_\alpha}{a_\beta} = \hbar^{\mu_\beta} q^{n + m}.    
\end{equation}
is satisfied for a fixed pair of $n, m \geq 0$ and some $\alpha \geq i$,  $\beta \leq i$, so in particular $\alpha \geq \beta$. Since $|\hbar^{\mu_\beta}| < 1$, $|q| < 1$, this means $|a_\alpha/a_\beta| < 1$ for some $\alpha > \beta$. It follows that the contour integral is nonsingular as long as $|a_\alpha| < |a_\beta|$ for $\alpha < \beta$, that is, in the chamber $|a_1| < \dots < |a_n|$ of $a$-variables. 
 \subsubsection{}
For $|z| < 1$, the integral may be evaluated explicitly by closing the contour to pick up sequences of poles accumulating at $|x| \to \infty$. The result may be expressed as 
\begin{equation}
\mathscr{F}\indices{^i_j}(a_\alpha; z) = \sum_{k \geq i} \mathscr{P}\indices{^i_k}(a_\alpha; z) \mathbf{V}\indices{^k_j}(a_\alpha; z). 
\end{equation}
The object $\mathbf{V}\indices{^k_j}$ is the vector vertex function, counting quasimaps approaching the fixed point $k$ at infinity with the insertion of the K-theoretic stable envelope $j$ at the origin. We have defined the matrix 
\begin{equation}
\mathscr{P}\indices{^i_k}(a_\alpha; z) := \mathscr{P}^i(x = a_k; a_\alpha; z)    
\end{equation}
known as the pole subtraction matrix, since it subtracts the poles of the $a$-variables from the vector vertex function ($z$-solution to qKZ) to form the $a$-solutions $\mathscr{F}\indices{^i_j}$---note the triangularity inherent in stable envelopes ensures it vanishes for $k < i$. The index $j$ is the one acted on by the $q$-difference connection, and the index $i$ labels which solution of qKZ one obtains. Taken together, the matrix $\mathscr{F}$ comprises the fundamental solution matrix of the qKZ equation, analytic in the chamber $|a_1| < \dots < |a_n|$. While the combinatorics of residues becomes increasingly complicated, general properties of the elliptic stable envelopes ensure that such pole cancellation always happens for $k > 1$ \cite{ao16}. 
\subsubsection{}
For each chamber $\mathfrak{C}$ of $a$-parameter space, there is a basis of stable envelopes and associated pole subtraction matrix $\mathscr{P}_{\mathfrak{C}}$ obtained essentially by evaluating them on fixed points. The fundamental solution matrix to qKZ in the chamber $\mathfrak{C}$ is obtained from the matrix of vector vertex functions $\mathbf{V}$ (with components $\mathbf{V}\indices{^k_j}$) simply by matrix multiplication: 
\begin{equation}
\mathscr{F}_{\mathfrak{C}} = \mathscr{P}_{\mathfrak{C}} \mathbf{V}.    
\end{equation}
Importantly, the vector vertex function itself is insensitive to the choice of chamber. The monodromy problem of qKZ is to understand the connection matrix relating two chambers $\mathscr{F}_{\mathfrak{C}}$ and $\mathscr{F}_{\mathfrak{C}'}$. The machinery of elliptic stable envelopes allows this problem to be solved in one line: 
\begin{equation}
\mathscr{F}_{\mathfrak{C}'} = \mathscr{P}_{\mathfrak{C}'} \mathbf{V} = \mathscr{P}_{\mathfrak{C}'} \mathscr{P}^{-1}_{\mathfrak{C}} \mathscr{P}_{\mathfrak{C}} \mathbf{V} = (\mathscr{P}_{\mathfrak{C}'} \mathscr{P}^{-1}_{\mathfrak{C}}) \mathscr{F}_{\mathfrak{C}}. 
\end{equation}
The monodromy matrix is then clearly, up to minor details, the change of basis matrix for elliptic stable envelopes. By definition, this is the geometric elliptic $R$-matrix. 

\subsection{Conformal Limit}
Consider now the conformal limit, 
$$q \to 1, \qquad \hbar = q^{-1/\kappa} \to 1,
$$
with the level $\kappa$ fixed. From the string theory perspective, which we return to shortly, this limit corresponds to the one taking the six dimensional $(2, 0)$ little string theory to the six dimensional $(2, 0)$ SCFT \cite{ah15, afo17}. One also scales the FI parameter as $\zeta = \eta \log (q)$, with $\eta$ fixed.

In this limit, 
$$\mathscr{U}_{\hbar}(\widehat{\mathfrak{sl}_2}) \rightarrow (\widehat{\mathfrak{sl}_2})_{k},$$
and the qKZ equation becomes the corresponding KZ equation. We will recover this directly by showing that the integral solutions of one reduce to integral solutions of the other.
\subsubsection{}

In this limit, one has 
\begin{equation}
\prod_{i \neq j} \frac{\varphi_q(x_i/x_j)}{\varphi_q(tx_i/x_j)} \to \prod_{i \neq j}(1 - x_i/x_j)^{1 + \frac{1}{\kappa}} 
\end{equation}
and likewise 
\begin{equation}
\prod_{i = 1}^k \prod_{\alpha = 1}^n \frac{\varphi_q(qx_i/m_\alpha^-)}{\varphi_q(x_i/m_\alpha^+)} \to \prod_{i = 1}^k \prod_{\alpha = 1}^n (1 - x_i/a_\alpha)^{-1 - \frac{\mu_\alpha}{\kappa}}.
\end{equation}
The stable envelopes become 
\begin{equation}
\Delta_\hbar(x)^{-1} \text{Stab}^K_\lambda(x) \to \prod_{i \neq j}(1 - x_i/x_j)^{-1} \prod_{i = 1}^k \prod_{\alpha = 1}^n (1 - x_i/a_\alpha) \sum_{\rho \in \Sigma_{\lambda}} \prod_{i = 1}^k\Big(1 - \frac{x_i}{a_{\rho(i)}}\Big)^{-1}
\end{equation}
and the sum runs over the set $\Sigma_\lambda$ of maps $\rho: \{1, \dots, k \} \to \{ 1, \dots, n \}$ in the orbit of $\rho(i) = \lambda_i$ under $S_k$ (it is what remains of the symmetrization). 

The integral \eqref{stabint} becomes 
\begin{equation}
\psi_\lambda(a_1, \dots, a_n) = \int [dx] \prod_{i = 1}^k x_i^\eta \prod_{i \neq j}(1 - x_i/x_j)^{1/\kappa} \prod_{i = 1}^k \prod_{\alpha = 1}^n(1 - x_i/a_\alpha)^{-\mu_\alpha/\kappa} \sum_{\rho \in \Sigma_\lambda} \prod_{i = 1}^k \Big(1 - \frac{x_i}{a_{\rho(i)}}\Big)^{-1}.
\end{equation}
In this way, we have rediscovered the integral representation of the solutions to Knizhnik-Zamolodchikov equations corresponding to the conformal blocks of $\widehat{\mathfrak{sl}}_2$ at level $k = \kappa - h^\vee$ on the cylinder, with Verma module vertex operators inserted at punctures (see \cite{kzbook} for a review of these).

\section{Generalization to Simply Laced $\mathfrak{g}$} \label{generalg}
In this section we describe a conjectural generalization of the structures studied here to general simply laced Lie algebras $\mathfrak{g}$. 

The proposal is based on the string theory origin, which we explain below.

\subsection{The Variety $X$ for general ${\mathfrak g}$}
The variety $X$ whose equivariant K-theory $K_T(X)$ should be identified with a subspace of fixed weight in a tensor product of $n$ Verma modules for $\mathscr{U}_\hbar(\widehat{\mathfrak{g}})$,
 can be described, conjecturally, as the moduli space of based quasimaps from $\mathbb{P}^1$ to a specific Nakajima variety $Y_n$ associated to the Dynkin diagram of $\mathfrak{g}$. 

$X$ is a partial compactification of the space of based rational maps from $\mathbb{P}^1 \to Y_n$. The degree of the quasimaps, which is now a vector with $\text{rk} (\mathfrak{g})$ components, specifies a certain weight space in the tensor product of $\mathscr{U}_\hbar(\widehat{\mathfrak{g}})$ Verma modules, which is taken to be fixed throughout in what follows. See figures 2-6 in \cite{ah15} for the choices of node and framing spaces describing $Y_1$---for $Y_n$, one should multiply the ranks of all spaces there by $n$.
The infinite-dimensionality of Verma modules is reflected in the fact that quasimaps may have arbitrarily high degree. For $\mathfrak{gl}_N$ with one puncture, one should again make contact with the results of \cite{tsymb09}.

\subsubsection{}
Because we consider only the most generic tame ramifications, that is, generic Verma module highest weights,  the variety $Y_n$ is fixed uniquely by the choice of $\mathfrak{g}$ and the number of punctures $n$ on the cylinder $\mathbb{C}^\times$. For less generic ramifications, which are discussed carefully in \cite{haouzi23}, other choices of $Y$ become relevant. In the special $\mathfrak{g} = \mathfrak{sl}_N$ case, $X$ is the moduli space of based quasimaps $\mathbb{P}^1 \to T^*(\text{partial flag variety})$, known as a parabolic Laumon space, and studied from the handsaw quiver perspective in \cite{nak11}. 

\subsubsection{}
Most of the parameters important for representation theory are identified as equivariant parameters, associated to automorphisms of the variety $X$.

Let $G_W \subset \text{Aut}(Y_n)$ act by changes of basis in the framing spaces, and let $G_F \subset G_W$ be the stabilizer of $f(\infty) \in Y_n$, for $f \in X$. $G_F$ has rank $n(1 + \text{rk}(\mathfrak{g}))$. The group 
\begin{equation}
G_F \times \mathbb{C}^\times_\hbar \subset \text{Aut}(X)
\end{equation}
acts on $X$. The action of the first factor on $Y_n$ induces an action on $X$, and the second factor arises from automorphisms of the domain curve of quasimaps. The equivariant parameters for the maximal torus $T \subset G_F \times \mathbb{C}^\times_\hbar$ are likewise identified with parameters of some $q$-deformed conformal blocks; the positions $a_\alpha$ of the punctures and Verma highest weights $\mu_\alpha$ come from the maximal torus in $G_F$, and the $\hbar$ parameter of quantum affine algebra is the equivariant parameter scaling the domain $\mathbb{P}^1$. 

Eventually we will study maps $\mathbb{P}^1 \to X$, and the equivariant parameter scaling the base $\mathbb{P}^1$ (this is the $\mathbb{P}^1$ going to $X$, not to be confused with the $\mathbb{P}^1$ going to $Y_n$) is the shift parameter $q$ of qKZ.

\subsubsection{}
$X$ may itself be presented as a subvariety of a GIT quotient given by the critical locus of a certain invariant function on the prequotient, called the superpotential. In other words, $X$ is a Higgs branch of a quiver gauge theory with $\mathscr{N} = 2$ supersymmetry. For $\mathfrak{g}$ of type $A$, this has been appreciated for some time and reduces to the construction of Nakajima \cite{nak11} given in the language of handsaw quiver varieties. For $\mathfrak{g}$ of $D$ and $E$ types, the analogous handsaw quivers are presented in section 6 of \cite{ah15}. In all cases, the superpotential is the standard $\mathscr{N} = 4$ superpotential associated to the nodes of the quiver, with an extra cubic term compatible with flavor symmetries for each ``tooth'' of the handsaw. 

\subsubsection{}
Since $X$ naturally embeds in a GIT quotient, it carries its own enumerative theory of quasimaps, and one can study its equivariant quantum K-theory, as in \cite{okounkov15}. The equivariant K-theoretic vertex function of $X$ is manifestly a $\mathscr{W}_{q, t}(\mathfrak{g})$ conformal block of the type described above (there are integral formulas for both, and it is elementary to check that they coincide). For $\mathfrak{g} = \mathfrak{sl}_2$, we essentially saw this explicitly in section \ref{quivergauge} and will recall it in more detail in section \ref{qlanglands}. For general simply laced $\mathfrak{g}$, this is a result of \cite{ah15}. 
\subsubsection{}
For quantum affine algebra blocks to arise from the geometry of $X$, we expect that a version of the stable envelope construction must exist for $X$. For $\mathfrak{g} \neq A_1$, $X$ does not quite fit the assumptions of \cite{okounkov21}, \cite{okounkov20} as stated, due to the presence of a superpotential. We expect that this is not a serious difficulty, and that existence may be established for all such $X$ described; the stable envelopes would need to be understood as certain classes in a critical $K$-theory with respect to the superpotential. 

We expect that the quasimap spaces $X$ are the natural setting to extend much of the theory explained in \cite{okounkov15} to Verma module representations, at least for the simply laced $\mathfrak{g}$.

\subsection{String Theory Origin}
Our proposal is motivated by string theory. The basic setup has been explained in detail in \cite{ah15}, so we shall be brief. 

The string theory underlying the constructions is the six dimensional $(2, 0)$ little string theory. In the limit of vanishing string size, the $(2, 0)$ little string reduces to the celebrated six dimensional $(2, 0)$ superconformal field theory (SCFT). This limit coincides with the $q \to 1$ limit which sends the deformed chiral algebras back to the ordinary chiral algebras.

\subsubsection{}
The little string theory is studied on the six-manifold 
\begin{equation}
\mathbb{C}_q \times \mathbb{C}_{t^{-1}} \times \mathcal{C}
\end{equation}
where $\mathcal{C}$ is the cylinder on which the conformal blocks will eventually live, and $q, t^{-1}$ are $\Omega$-background parameters defined below.

The little string theory of type $\mathfrak{g}$, for $\mathfrak{g}$ a simply laced Lie algebra, arises as a decoupling limit of IIB string theory on 
\begin{equation}
\mathbb{C}_q \times \mathbb{C}_{t^{-1}} \times \mathcal{C} \times \mathscr{X}_\mathfrak{g}
\end{equation}
where $\mathscr{X}_\mathfrak{g}$ is a resolved ALE surface of type $\mathfrak{g}$. The string coupling of IIB is sent to zero in such a way that the ten-dimensional bulk dynamics is decoupled, and what is left is a six-dimensional theory supported near the exceptional divisor of $\mathscr{X}_\mathfrak{g}$.
\subsubsection{}
Introducing punctures supporting Verma modules means introducing codimension two defects of the six dimensional little string supported near the points of $\mathcal{C}$ called $a_\alpha$ throughout. These uplift to D5 branes of the full IIB theory, wrapping $\mathbb{C}_q \times \mathbb{C}_{t^{-1}}$ and noncompact 2-cycles in $\mathscr{X}_\mathfrak{g}$. The D5 branes come in clusters, each corresponding to a vertex operator. For a Verma module of a generic weight,  the number of D5 branes in a cluster is $\text{rk}(\mathfrak{g}) + 1$. The positions of D5 branes in such a cluster on $\mathcal{C}$ specify both $a_\alpha$ and the highest weight $\mu_\alpha \in \mathfrak{h}^*$ associated to the puncture. 

In addition to these D5 branes, one introduces D3 branes, corresponding to codimension four dynamical defects of the little string, supported on $\mathbb{C}_q$ as well as compact 2-cycles in $\mathscr{X}_\mathfrak{g}$. The compact 2-cycles in $\mathscr{X}_\mathfrak{g}$ correspond to simple roots of $\mathfrak{g}$ upon identifying $H_2(\mathscr{X}_\mathfrak{g}; \mathbb{Z})$ with the root lattice of $\mathfrak{g}$, and are related to a choice of fixed weight subspace as in the mathematical constructions explained above.
\subsubsection{} 
To obtain the conformal blocks of $\mathscr{U}_\hbar(\widehat{\mathfrak{g}})$ or $\mathscr{W}_{q, t}(\mathfrak{g})$, one computes the partition function of little string theory in this background. This is essentially a supersymmetric index:
\begin{equation}
Z = \Tr (-1)^F q^{S - S_H} t^{S_H - S_V}
\end{equation}
where $S$ generates the rotation of one of the complex planes in $\mathbb{C}^2$, $S_V$ rotates the other plane, and $S_H$ is an R-symmetry generator whose inclusion is required to preserve supersymmetry. Strictly speaking, one should also include a flavor symmetry insertion in the trace, but we suppress this to lighten the notation. $q$ and $t^{-1}$ are the equivariant parameters for such rotations, and correspond to Nekrasov's $\Omega$-background parameters. The trace is a trace going around the T-dual circle to the circle in $\mathcal{C}$. 
\subsubsection{}
Due to supersymmetry, $Z$ receives contributions only from modes supported on the defects, which are the modes supported on the noncompact part of the D3 brane worldvolume. The D5 branes wrapping noncompact cycles are much heavier than these D3 branes, therefore they are treated as essentially classical and their only role is to contribute the sector of matter content corresponding to the D3-D5 strings to the worldvolume theory of D3 branes. This reduces the computation of $Z$ to the computation of the $\Omega$-background partition function of a 3d gauge theory. The theory is three-dimensional rather than two dimensional because the circle in $\mathcal{C}$ transverse to the D3 brane worldvolume contributes string winding modes; this circle can be made geometric in the gauge theory by applying T-duality, after which winding modes become Kaluza-Klein modes of the 3d theory on the dual circle. 
\subsubsection{}
The three-dimensional gauge theory whose Higgs branch is $X$ is the theory on the worldvolume of the D3 branes. It has $\mathscr{N} = 2$ supersymmetry in three dimensions. The Nakajima variety $Y_n$ is the Higgs branch of the theory on D5 branes, in which D3 branes appear as vortices. The D5 branes being heavy enough to be considered classical is the statement that the five-dimensional gauge theory is on its Higgs branch, so the BPS objects it supports are the vortices, which are the D3 branes. All interesting dynamics happens on the vortices.

The 3d partition function $Z$ is exactly the vertex function of the Higgs branch $X$, and coincides with a $\mathscr{W}_{q, t}(\mathfrak{g})$ conformal block. To produce $\mathscr{U}_\hbar(\widehat{\mathfrak{g}})$ blocks, one introduces supersymmetric line defects in the 3d gauge theory corresponding to K-theoretic stable envelopes of $X$, which uplift to orbifold defects of the full six-dimensional little string. 

\subsection{Less Generic Ramifications}
While this paper focuses on the case of tame ramifications (meaning generic Verma module highest weights), the string theory construction described above predicts the statement of the correspondence for the less generic ramification types. To describe this, it is important to recall more details of the construction of \cite{ah15}, generalized and elaborated on in \cite{haouzi2017}. Since the aim of the discussion is to discuss the ramification data at a given puncture, it suffices to consider only a single puncture for simplicity. 
\subsubsection{}
When the five-dimensional gauge theory is on its Higgs branch, the D5 branes of the IIB string wrap a collection of non-compact cycles in $\mathscr{X}_{\mathfrak{g}}$. By a classical result, the relative homology $H_2(\mathscr{X}_\mathfrak{g}, \partial \mathscr{X}_\mathfrak{g}; \mathbb{Z})$ is identified with the weight lattice $\Lambda_W$ of $\mathfrak{g}$. The homology classes of the D5 branes inside of $\mathscr{X}_\mathfrak{g}$ define a collection of weights 
\begin{equation}
\omega_i = [S^*_i] \in \Lambda_W.
\end{equation}
The allowed collections of weights $\{ \omega_i \}$ classify the codimension two defects of the $(2, 0)$ little string---this collection of weights describes precisely the group of D5 branes which describe the ramification data at a single puncture of $\mathcal{C}$ (in the conformal limit, the branes collide, so that one has a single puncture on $\mathcal{C}$, but before the limit the puncture is identified with the center of mass position of the group of D5 branes). Different choices of defect correspond to different choices of ramification data at punctures. The choices of weights are restricted by the condition that each $\omega_i$ must lie in a fundamental representation of $\mathfrak{g}$, and that 
\begin{equation}
\sum_i \omega_i = 0.
\end{equation}
The latter is just the condition that the net D5 brane flux vanishes; it is equivalent to the condition that the 5d gauge theory on the D5 worldvolume is asymptotically conformal. 
\subsubsection{}
The most canonical type of defect, describing the tame ramifications discussed above, corresponds to satisfying the above conditions by taking $\text{rk}(\mathfrak{g}) + 1$ such weights which sum to zero and span the weight lattice---more details and explicit examples of such weight systems for all ADE Lie algebras may be found in \cite{ah15}. In the conformal limit, these describe Toda primaries with generic complex momenta, ``generic'' meaning precisely that the corresponding Verma module is free of null states. 

However, it is possible to tune the highest weights of Verma modules such that null states appear at levels below the highest weight (this happens e.g. when the highest weight lies on the boundary of a Weyl chamber). In the string theory construction, this merely corresponds to other choices of weight systems $\{ \omega_i \}$, with less than $\text{rk}(\mathfrak{g}) + 1$ weights. Remarkably, one can use this picture to construct explicitly the corresponding $\mathscr{W}_{q, t}(\mathfrak{g})$ vertex operators, which become non-generic primaries in the conformal limit. 

When one studies this construction carefully, one finds that such vertex operators are naturally constructed using the representation theory of $\mathscr{U}_\hbar(\widehat{\mathfrak{g}})$, in agreement with the ramified quantum $q$-Langlands program. This will be explained in much greater detail in a work of Haouzi \cite{haouzi23}.
\subsubsection{}
On the geometric side, the effect of considering less generic ramifications is to change the Nakajima variety $Y_n$ that appears as the Higgs branch of the gauge theory on the D5 branes. For the other ramifications, one will find other Nakajima varieties based on the Dynkin diagram of $\mathfrak{g}$, with the dimension vectors for the node and framing spaces encoded in the choice of weight system $\{ \omega_i \}$ describing the puncture--see \cite{haouzi2017} for more details as well as some examples. The rest of the story proceeds as before: one constructs the moduli space $X$ of quasimaps to this Nakajima variety, and computes its K-theoretic vertex function. $X$ admits a handsaw quiver-type description, which may also be read off from the weight system $\{ \omega_i \}$ specifying the choice of defect. 

The K-theoretic vertex function of $X$ will coincide with the $\mathscr{W}_{q, t}(\mathfrak{g})$ conformal block, with non-generic vertex operators. Presumably, stable envelopes will exist for this more broad class of varieties, and one may use them to produce the $\mathscr{U}_\hbar(\widehat{\mathfrak{g}})$ conformal blocks with non-generic Verma module representations. 

\subsection{AGT Correspondence and Gauge/Vortex Duality}
The variety $X$ we have studied throughout this paper, and the generalizations discussed above, have an interpretation as a moduli space of rational curves. To find the qKZ equation, one studies enumerative geometry of curves in $X$ itself. This naturally suggests that the problem has a origin in complex dimension two, and this is indeed the case. 

\subsubsection{}
It follows from the little string origin of the 3d $\mathscr{N} = 2$ theory, that the vertex function of $X$ coincides with the Nekrasov partition function of the 5d $\mathscr{N} = 1$ supersymmetric  gauge theory whose Higgs branch is $Y_n$. In the ${\mathfrak{g}} = {\mathfrak{su}_2}$ case, the theory has gauge group $U(n)$ and $2n$ fundamental hypermultiplets.

The computations we perform in this paper in three-dimensional theories can be thought of as capturing the behavior of this five-dimensional gauge theory on its Higgs branch. This correspondence between 5d and 3d dynamics is sometimes called gauge/vortex duality, see \cite{ah15} for a review. It is not unlike the relation between counts of quasimaps in Hilbert schemes of points and certain Donaldson-Thomas counts, see \cite{okounkov15}.  
\subsubsection{}
An important source of inspiration for the present work were the results of \cite{Jeong_2021}, \cite{Lee_2021}, \cite{nek22} on KZ equation and spin chains from four-dimensional (cohomological) instanton counting. Our results are in some sense both a restriction and a generalization. 

They are a restriction in the sense that we limit the considerations to the Higgs branch of the bulk 5d gauge theory, so do not see the more exotic Heisenberg-Weyl modules that appeared in \cite{Jeong_2021}, \cite{Lee_2021}, \cite{nek22} at generic values of Coulomb moduli. They are a generalization in the sense that our techniques work in K-theory, and take a step towards placing this class of results in the general framework of stable envelopes. It would be interesting to further develop the relationship of the qKZ equation studied here to the of KZ equation of \cite{nek22} and its generalizations to other $A$-type quiver gauge theories.

\section{Application: Langlands with Ramifications} \label{qlanglands}
The $q$-deformed quantum geometric Langlands correspondence proposed and proved for simply laced $\mathfrak{g}$ in \cite{afo17} originated from the study of three-dimensional gauge theories with $\mathscr{N} = 4$ supersymmetry, softly broken to $\mathscr{N} = 2$ by a particular twisted mass. This correspondence concerned the conformal blocks with operators labeled by finite-dimensional, in fact, fundamental $\mathfrak{g}$-modules at the punctures. 

The generalization of this structure to Langlands with ramifications involves replacing the vertex operators at punctures with those of Verma module type, which are precisely the kind under consideration in this paper. The ramified quantum $q$-Langlands correspondence is a conjectural explicit correspondence of $q$-conformal blocks of $\mathscr{U}_\hbar(\widehat{^L \mathfrak{g}})$ and $\mathscr{W}_{q, t}(\mathfrak{g})$, of the same type as \cite{afo17}: 
\begin{equation}
\text{specific covector} \times \text{$\mathscr{U}_\hbar(\widehat{^L \mathfrak{g}})$ block} = \text{$\mathscr{W}_{q, t}(\mathfrak{g})$ block} 
\end{equation}
but with a different choice of vertex operators. The parameters of $\mathscr{U}_\hbar(\widehat{^L \mathfrak{g}})$ and $\mathscr{W}_{q, t}(\mathfrak{g})$ are identified as  
\begin{equation}
t = q/\hbar. 
\end{equation}
One one side of the correspondence, we have $n$ Verma type vertex operators of quantum affine algebra with highest weights $\mu_\alpha$, and on the other side we have $q$-deformed $\mathfrak{g}$-Toda primaries of momenta $\gamma_\alpha$, $\alpha = 1, \dots, n$. In the rank one $\mathfrak{g} = \mathfrak{sl}_2$ case that we study here, these are $n$-tuples of complex numbers related by 
\begin{equation}
\gamma_\alpha = \frac{\mu_\alpha}{\kappa} + 1. 
\end{equation}
The highest weight of the Verma module at a given puncture determines the ramification data there. The genericity of this highest weight determines the ramification type of the puncture. 

We direct readers to the paper \cite{haouzi23} for a much more general and thorough overview of the ramified quantum $q$-Langlands program and its origin in the physics of little string theory and three-dimensional gauge theories. In this short section our goal is just to explain the application of the stable envelopes and geometric considerations of this paper to this context.

\subsubsection{}
The modern understanding of the geometric Langlands program, for a complex reductive group $G_{\mathbb{C}}$ and Riemann surface $\cal C$, is as an equivalence of categories of $\mathscr{D}$-modules on the moduli stack of $^L G_{\mathbb{C}}$-bundles and coherent sheaves on the space of $G_{\mathbb{C}}$-local systems on $\cal C$. The reason why the explicit correspondences of conformal blocks discussed above makes contact with the general geometric Langlands program goes back to the original constructions of Beilinson and Drinfeld. In particular, they made use of a canonical identification of the center of the universal enveloping algebra of $\widehat{^L \mathfrak{g}}$ at the critical level $\kappa = 0$ with the classical $\mathscr{W}$-algebra $\mathscr{W}_\infty(\mathfrak{g})$. The so-called Hecke eigensheaves that play a role in the Langlands correspondence may arise as sheaves of conformal blocks of $\widehat{^L \mathfrak{g}}$ at the critical level; likewise the opers on the other side of Langlands duality can be understood as a semiclassical limit of Liouville/Toda conformal blocks. 

It is thus natural from the conformal field theory point of view to deform away from the critical level, leading to the so-called quantum geometric Langlands correspondence. Working moreover with the quantum affine algebra and $(q, t)$-deformed $\mathscr{W}$-algebra corresponds to a double deformation of the original Langlands correspondence: moving away from the critical level, and passing from conformal blocks to $q$-deformed conformal blocks. The important point for the present work is that in the Langlands program, certain ramifications correspond to inserting vertex operators of Verma module type; see \cite{frenkel05} for a review of the connections between the Langlands program and conformal field theory. 

\subsubsection{}
In this work, we  have identified in the geometry of $X(k, n)$ the qKZ equations for $\mathscr{U}_\hbar(\widehat{\mathfrak{sl}_2})$. On the other hand, as reviewed in \cite{afo17}, \cite{ah15} the Mellin-Barnes integrals for vertex functions are often manifestly equal to the free-field integral formulas for conformal blocks of a deformed $\mathscr{W}$-algebra, in this case $\mathscr{W}_{q, t}(\mathfrak{sl}_2)$. This setup allows for one to construct an explicit correspondence of basis vectors in the spaces of conformal blocks, of the type 
\begin{equation}
\text{specific covector} \times \text{$\mathscr{U}_\hbar(\widehat{\mathfrak{sl}_2})$ block} = \text{$\mathscr{W}_{q, t}(\mathfrak{sl}_2)$ block}.  
\end{equation}
The basic idea underlying this is simple: the $\mathscr{W}_{q, t}(\mathfrak{sl}_2)$ block comes from the partition function of the 3d $\mathscr{N} = 2$ theory on $\mathbb{C} \times S^1$ with no insertion at $0 \in \mathbb{C}$, while the $\mathscr{U}_\hbar(\widehat{\mathfrak{sl}_2})$ block is a vector, with components given by insertion of stable envelopes at $0 \in \mathbb{C}$. Since stable envelopes form a basis in $K_T(X)$, one can expand the trivial insertion $1 = \mathscr{O}_X \in K_T(X)$ in this basis, which gives rise to the explicit isomorphism above. 

\subsubsection{}
Let us recall in some more detail the connection to deformed $\mathscr{W}$-algebras developed in \cite{ah15, afo17}. The basic observation is that the integral formula for the partition function of the 3d $\mathscr{N} = 2$ theory is simply manifestly equal to an expression for a ($q$-)conformal block of $\mathscr{W}_{q, t}(\mathfrak{sl}_2)$ algebra, in the free field formalism. The free field realization of the $\mathscr{W}_{q, t}(\mathfrak{g})$ algebra, for more general $\mathfrak{g}$, is reviewed in section 2.2 of \cite{afo17} as well as the appendix of \cite{ah15}. 

The $q$-conformal blocks are correlators of $n$ primary vertex operators of $\mathscr{W}_{q, t}(\mathfrak{sl}_2)$ at generic complex momenta $\gamma_\alpha$ inserted at punctures $a_\alpha \in \cal A \simeq \mathbb{C}^\times$, denoted $V^\vee_{\gamma_\alpha}(a_\alpha)$: 
\begin{equation}
\bra{\mu'} V^\vee_{\gamma_1}(a_1) \dots V^\vee_{\gamma_n}(a_n) (Q^\vee)^k \ket{\mu}, 
\end{equation}
in the presence of screening charges
\begin{equation}
Q^\vee = \int dx S^\vee(x).
\end{equation}
The explicit expressions for screening currents $S^\vee(x)$ are given in \cite{frenkelreshetikhin}. The primary operators at generic complex momenta were constructed in appendix of \cite{ah15}, generalizing degenerate vertex operators given in \cite{frenkelreshetikhin}. $\ket{\mu}$ denotes another generic primary, with momentum $\mu$.
For this to not trivially vanish by momentum conservation, we must have $\mu' - \mu = 2k$. 

Since the operators are constructed from free fields, their correlation functions may be evaluated by pairing and contracting. Using \cite{frenkelreshetikhin}
\begin{equation}
\prod_{i < j} \bra{0} S^\vee(x_i) S^\vee(x_j) \ket{0} =  \prod_{i \neq j} \frac{\varphi_q(x_i/x_j)}{\varphi_q(tx_i/x_j)} \times \prod_{i < j} \frac{\theta(tx_i/x_j)}{\theta(x_i/x_j)}
\end{equation}
where $\theta(x) = \varphi_q(x)\varphi_q(q/x)$, and \cite{ah15}
\begin{equation}
\bra{0} S^\vee(x_i) V^\vee_{\gamma_\alpha}(a_\alpha) \ket{0} = \frac{\varphi_q(q^{\gamma_\alpha} x_i/a_\alpha)}{\varphi_q(x_i/a_\alpha)}
\end{equation}
one finds that the integral agrees with the integral computing the partition function, up to a product of theta functions and provided parameters are identified via 
\begin{equation}\label{qV}
a_\alpha = m_\alpha^+, \qquad  q^{\gamma_\alpha} = q  m_\alpha^+/ m_\alpha^-,
\end{equation}
and $\mu = \zeta + \text{const}$. The product of theta functions turns out to be inessential, as it just contributes this constant shift (see discussion in section 3.5 of \cite{afo17} and below). In this way, the $q$-conformal block of the algebra $\mathscr{W}_{q, t}(\mathfrak{sl}_2)$ is identified with the vertex function of the variety $X$.

\subsection{Case Study: Abelian Theory and Langlands Correspondence}
In the case $k = 1$, the gauge group is abelian and the above remarks may be made completely explicit and turned into a proof. Both the geometry of the Higgs branch $X$ and the integral formulas drastically simplify.
\subsubsection{}
The Higgs branch $X$ is in this case a toric variety, and a rather simple one:
\begin{equation}
X = \mathscr{O}(0) \oplus \mathscr{O}(-1)^{\oplus n} \to \mathbb{P}^{n - 1}.
\end{equation}
Explicitly, this space is presented as the quotient of the stable locus of $(B, I_1, \dots, I_n, J_1,\dots, J_n) \in \mathbb{C}^{2n + 1}$ by $\mathbb{C}^\times$, where stability simply means that all $I_\alpha$ do not simultaneously vanish. $\mathbb{C}^\times$ acts trivially on $B$, scales $I_\alpha$ with weight $+1$, and scales $J_\alpha$ with weight $-1$. 
\subsubsection{}
The torus $T$ acts on $X$ with $n$ isolated fixed points, labeled by an index $i = 1, \dots, n$. The corresponding stable envelopes are given by 
\begin{equation}
\text{Stab}^K_i(x) = \prod_{\alpha = 1}^{i - 1} \Big(1 - \frac{x}{a_\alpha}\Big) \prod_{\alpha = i + 1}^n \Big( 1 - \frac{x}{\hbar^{\mu_\alpha} a_\alpha} \Big) = \pm(\text{line bundle}) \times \mathscr{O}_{Z_i} \in K_T(X). 
\end{equation}
$\mathscr{O}_{Z_i}$ denotes the K-class of the structure sheaf of the subvariety $Z_i \subset X$ defined by the equations
\begin{equation*}
Z_i = \{ I_1 = \dots = I_{i - 1} = J_{i + 1} = \dots = J_n = 0 \} \subset X.    
\end{equation*}
Each $Z_i$ is a codimension $n - 1$ holomorphic cycle (B-brane) inside of $X$. Geometrically, $Z_i$ is the attracting locus of the $i$-th component of the $A$-fixed locus with respect to the chamber of real mass parameters used to define the stable envelope. 

In this case, the integral \eqref{stabint} becomes 
\begin{equation}
\psi_i (a_1, \dots, a_n) = \int_\gamma \frac{dx}{2 \pi ix} x^{-\frac{\log(z)}{\log q}} \prod_{\alpha = 1}^n \frac{\varphi_q(\hbar^{-\mu_\alpha}x/a_\alpha)}{\varphi_q(x/a_\alpha)} \times \frac{1}{1 - \hbar^{-\mu_i}x/a_i} \prod_{\alpha = 1}^{i - 1} \frac{1 - x/a_\alpha}{1 - \hbar^{-\mu_\alpha}x/a_\alpha}.
\end{equation}
Up to a shift $a_\alpha \to \hbar^{-\mu_\alpha} a_\alpha$, this coincides with eq. 5.7 of \cite{afo17}, which gives the level one solutions to qKZ with generic highest weights; the Verma modules produce the analytic continuation of this formula to the complex-valued highest weights.
\subsubsection{}
We explained above that stable envelopes allow one to define an $R$-matrix completely geometrically, as the change of basis matrix between the stable bases in neighboring chambers of real mass parameters. In this section, we will show that this geometric definition recovers standard formulas for $R$-matrices in tensor products of $\mathscr{U}_\hbar(\widehat{\mathfrak{sl}_2})$ Verma modules. Restrict further to the case $n = 2$ for simplicity, so $X$ is a copy of $\mathbb{C}$ times the resolved conifold. 

The two chambers one considers in this case are the $(+)$ chamber $|a_1| < |a_2|$, and the $(-)$ chamber $|a_2| < |a_1|$. The stable envelopes in each chamber are (with minor rescalings by powers of $\hbar$, for convenience)
\begin{equation}
\begin{split}
\Phi_1^{(+)}(x) & = \hbar^{\mu_2/2}(1 - \hbar^{-\mu_2} x/a_2) \\
\Phi_2^{(+)}(x) & = 1 - x/a_1 \\
\Phi_1^{(-)}(x) & = 1 - x/a_2 \\
\Phi_2^{(-)}(x) & = \hbar^{\mu_1/2}(1 - \hbar^{-\mu_1}x/a_1).
\end{split}
\end{equation}
The change of basis matrix is computed from 
\begin{equation}
\begin{split}
\Phi^{(-)}_1(x) & = R_{11}\Phi_1^{(+)}(x) + R_{12} \Phi_2^{(+)}(x) \\
\Phi^{(-)}_2(x) & = R_{21}\Phi_1^{(+)}(x) + R_{22} \Phi_{2}^{(+)}(x).
\end{split}
\end{equation}
One easily finds
\begin{equation}
\begin{split}
R_{11} & = \frac{1 - a_1/a_2}{\hbar^{\mu_2/2} - \hbar^{-\mu_2/2}a_1/a_2} \\
R_{12} & = \frac{a_1}{a_2} \Big( \frac{\hbar^{-\mu_2} - 1}{\hbar^{-\mu_2}a_1/a_2 - 1} \Big) \\
R_{21} & = \frac{\hbar^{(\mu_1 - \mu_2)/2} - \hbar^{-(\mu_1 + \mu_2)/2}}{1 - \hbar^{-\mu_2}a_1/a_2} \\
R_{22} & = \frac{\hbar^{-\mu_1/2} - \hbar^{\frac{\mu_1}{2} - \mu_2} a_1/a_2}{1 - \hbar^{-\mu_2}a_1/a_2}.
\end{split}
\end{equation}
Setting $a_1/a_2 = \hbar^{-(\mu_1 - \mu_2)/2} a$, one recognizes in these formulas the expected $R$-matrix acting in the first weight subspace below the highest (compare eq. 5.2-5.4 in \cite{afo17}).
\subsubsection{}
Returning to the case of $k = 1$ for general $n$, it is manifest that the stable basis $\text{Stab}_i^K$ is triangular with respect to restrictions to fixed points. Since $1 = \mathscr{O}_X \in K_T(X)$, there is an expansion
\begin{equation}
1 = \sum_{i = 1}^n W_i \text{Stab}^K_i(x)
\end{equation}
and one can use the triangularity of the stable basis to solve for $W_i$'s recursively: setting $x = a_1$ determines $W_1$, setting $x = a_2$ determines $W_2$ given $W_1$, and so on. The existence of this expansion establishes the ramified quantum $q$-Langlands correspondence in the simplest case of $q$-conformal blocks for $\mathfrak{sl}_2$ at $k = 1$. 
\subsubsection{}
In establishing the quantum $q$-Langlands correspondence in general, it becomes increasingly inefficient to work directly with integral formulas, especially for $\mathfrak{g} \neq \mathfrak{sl}_2$. It is expected that the geometric techniques used in \cite{afo17} would extend straightforwardly to the present setting and establish the correspondence for simply-laced $\mathfrak{g}$, though we do not consider that direction in this paper. 

\newpage

\printbibliography

\end{document}